\documentclass[acmlarge, screen, nonacm]{acmart} 
\usepackage{adjustbox}
\usepackage[graphicx]{realboxes}
\usepackage{array, multirow, graphicx}
\usepackage{subcaption}
\usepackage{graphicx}
\usepackage{wrapfig}
\usepackage{lipsum}
\usepackage{tikz}
\usepackage{ragged2e}
\usepackage[font=normalsize,labelfont=bf]{caption}

\usepackage{adjustbox}
\usetikzlibrary{fadings}

\AtBeginDocument{%
  \providecommand\BibTeX{{%
    \normalfont B\kern-0.5em{\scshape i\kern-0.25em b}\kern-0.8em\TeX}}}

\newcommand{\msgtype}[1]{\textsc{#1}}
\newcommand{\msg}[1]{$\langle$\texttt{#1}$\rangle$}

\begin{document}

\title{Reaching Consensus in the Byzantine Empire: A Comprehensive Review of BFT Consensus Algorithms}

\author{Gengrui Zhang}
\affiliation{%
  \institution{University of Toronto}
  \city{Toronto}
  \country{Canada}}
\author{Fei Pan}
\affiliation{%
  \institution{University of Toronto}
  \city{Toronto}
  \country{Canada}}
\author{Yunhao Mao}
\affiliation{%
  \institution{University of Toronto}
  \city{Toronto}
  \country{Canada}}
\author{Sofia Tijanic}
\affiliation{%
  \institution{University of Toronto}
  \city{Toronto}
  \country{Canada}}
\author{Michael Dang'ana}
\affiliation{%
  \institution{University of Toronto}
  \city{Toronto}
  \country{Canada}}
\author{Shashank Motepalli}
\affiliation{%
  \institution{University of Toronto}
  \city{Toronto}
  \country{Canada}}
\author{Shiquan Zhang}
\affiliation{%
  \institution{University of Toronto}
  \city{Toronto}
  \country{Canada}}
\author{Hans-Arno Jacobsen}
\affiliation{%
  \institution{University of Toronto}
  \city{Toronto}
  \country{Canada}}
  
\email{gengrui.zhang@mail.utoronto.ca}
\email{fei.pan@mail.utoronto.ca}
\email{yunhao.mao@mail.utoronto.ca}
\email{sofia.tijanic@mail.utoronto.ca}
\email{michael.dangana@mail.utoronto.ca}
\email{shashank.motepalli@mail.utoronto.ca}
\email{shiquan.zhang@mail.utoronto.ca}
\email{jacobsen@eecg.toronto.edu}

\renewcommand{\shortauthors}{Gengrui Zhang et al.}

\begin{abstract}
Byzantine fault-tolerant (BFT) consensus algorithms are at the core of providing safety and liveness guarantees for distributed systems that must operate in the presence of arbitrary failures. Recently, numerous new BFT algorithms have been proposed, not least due to the traction blockchain technologies have garnered in the search for consensus solutions that offer high throughput, low latency, and robust system designs. In this paper, we conduct a systematic survey of selected and distinguished BFT algorithms that have received extensive attention in academia and industry alike. We perform a qualitative comparison among all algorithms we review considering message and time complexities. Furthermore, we provide a comprehensive, step-by-step description of each surveyed algorithm by decomposing them into constituent subprotocols with intuitive figures to illustrate the message-passing pattern. We also elaborate on the strengths and weaknesses of each algorithm compared to the other state-of-the-art approaches.
\end{abstract}



\maketitle


\section{Introduction}
Byzantine fault-tolerant (BFT) consensus algorithms, which coordinate server actions under Byzantine (arbitrary) failures, have been extensively studied due to the burgeoning development of blockchain applications~\cite{lamport1982byzantine, lamport1983weak}.
Since BFT algorithms tolerate arbitrary faults (i.e., the behavior of faulty servers (processes) is not constrained~\cite{schneider1990implementing}), they have long been used in safety-critical systems (e.g., aircraft~\cite{ wensley1978sift, siewiorek2005fault} and submarines~\cite{walter2005reliable}) where hardware may become unreliable in hostile environments (e.g., extreme weather and radiation). 
Recently, driven by the rising interest in blockchain technology, BFT algorithms have been widely deployed in numerous blockchain platforms to provide decentralized solutions that engender trust without relying on a third party, supporting cryptocurrencies~\cite{nakamoto2008bitcoin}, supply chains~\cite{queiroz2019blockchain}, international trade platforms~\cite{chang2020blockchain}, and the Internet of Things (IoT)~\cite{dorri2017towards}. In these applications, BFT algorithms are essential in providing correctness guarantees for consensus, as malicious behavior is becoming the norm~\cite{diem, gupta13resilientdb, gupta2021rcc, androulaki2018hyperledger}.

\begin{figure}
    \centering
    \begin{adjustbox}{max width=0.9\textwidth}
    \begin{tikzpicture}
        \coordinate (center) at (0,0);
        \tikzfading[name=arrowfading, top color=red!0, bottom color=red!100]
        
        \draw[->, red, line width=10pt] (0,0) -- (-30: 3cm) node[pos=1, below right]{\bf More robust BFT (\S\ref{sec:robustness})};
        
        \draw[->, blue, line width=10pt] (center) -- (90: 3cm) node[pos=1, above]{\bf More efficient BFT (\S\ref{sec:efficiency})};
        
        \definecolor{armygreen}{rgb}{0.29, 0.33, 0.13}
        \draw[->, armygreen, line width=10pt] (center) -- (-150: 3cm) node[pos=1, below left]{\bf More available BFT (\S\ref{sec:availability})};
        
        \fill[white] (0,0) circle (1cm);
        
        \node[align=center] at (0, 0) 
        {\textbf{Traditional BFT:}\\PBFT~\cite{castro1999practical}};

    \node[align=left] at (-4.7, 0) {
          \textbf{Multi-leader Consensus:}\\
          (e.g., MirBFT~\cite{stathakopoulou2019mir},
          ISS~\cite{stathakopoulou2022state})}; 

    \node[align=left] at (-4, 2.5) {
          \textbf{DAG-based protocols:}\\(e.g., DAG-Rider~\cite{keidar2021all}, Narwhal~\cite{danezis2022narwhal},\\
          Bullshark~\cite{spiegelman2022bullshark}, Shoal~\cite{spiegelman2023shoal})}; 
          
    \node[align=left] at (4.2, 2.5) {
          \textbf{Reducing message complexity:}\\
          (e.g., Zyzzyva~\cite{kotla2007zyzzyva},\\
          SBFT~\cite{gueta2019sbft}, HotStuff~\cite{yin2019hotstuff})}; 

    \node[align=left] at (4.5, -3.3) 
          {\textbf{Defending various attacks}\\ 
          (e.g., timing attacks: Prime~\cite{prime2011},\\ monopoly attacks: Spin~\cite{spin2009}, \\
          performance attacks: Aardvark~\cite{clement2009making})}; 

    \node[align=left] at (4.5, 1.2) 
          {\textbf{Ordering fairness:} \\
          (e.g., Pompe~\cite{zhang2020byzantine}, Fairy~\cite{stathakopoulou2021adding})};
          
    \node[align=left] at (-4., -3) {
          \textbf{Asynchronous and leaderless protocols:}\\
          (e.g., DBFT~\cite{crain2018dbft}, BEATS~\cite{duan2018beat},\\
          Honeybadger~\cite{miller2016honey})}; 

    \node[align=left] at (4.8, -0.5) 
          {\textbf{Reputation consensus:}\\(e.g., Prosecutor~\cite{zhang2021prosecutor}, \\
          PrestigeBFT~\cite{zhang2023prestigebft})};    
          
    \end{tikzpicture}
    \end{adjustbox}
    \caption{The taxonomy of state-of-the-art BFT consensus algorithms, where the position of each algorithm indicates its inclination toward one of the three categories.}
    \label{fig:taxonomyofbft}
\end{figure}
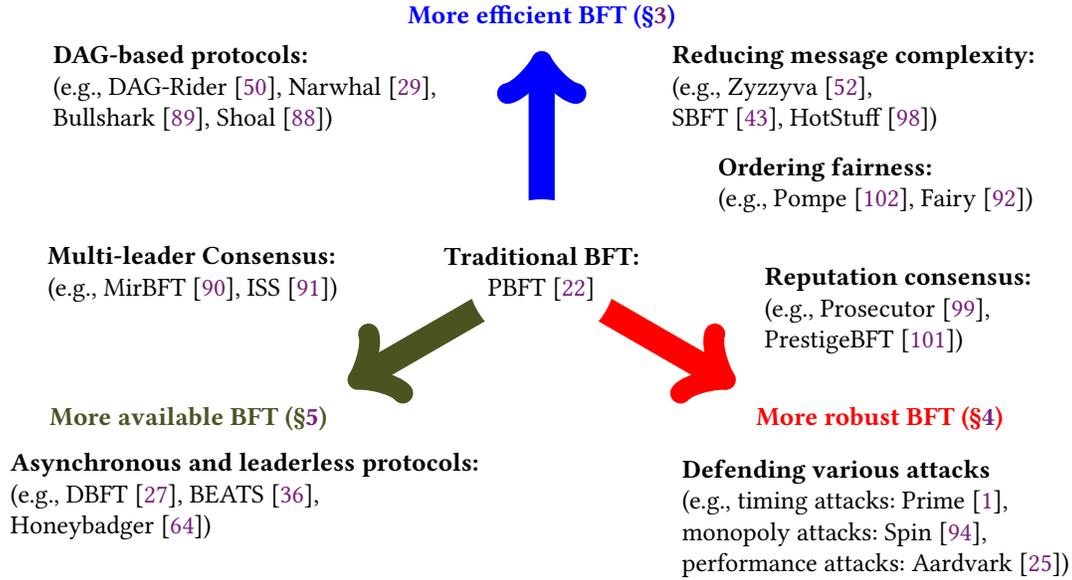

In this paper, we present a novel taxonomy for classifying state-of-the-art BFT consensus algorithms. We conduct a systematic survey of selected algorithms that have had significant impacts in both academia and industry. Our taxonomy presents three major categories: more efficient, more robust, and more available BFT approaches, each with subcategories representing distinct characteristics that contribute to the advancements in BFT consensus (shown in Figure~\ref{fig:taxonomyofbft}). We have also compared the message and time complexities of the surveyed algorithms in Table~\ref{tab:comparison-all-bft}. 

\begin{itemize}
    \item \textbf{More efficient BFT:} The first category includes BFT algorithms focused on improving replication efficiency, such as reducing message complexity and pipelining mechanisms before consensus.

    \item \textbf{More robust BFT:} The second category comprises BFT algorithms that prioritize robustness and security against various attacks and malicious behavior. Key aspects under this category include defending against various attacks, penalizing misbehavior, and improving fairness in ordering transactions.

    \item \textbf{More available BFT:} The third category incorporates BFT algorithms that prioritize availability and system responsiveness under different network conditions, including asynchronous protocols and leaderless protocols.

\end{itemize}

Due to the many optimizations of existing algorithms and the emergence of newly proposed algorithms, the effort required to understand these algorithms has been growing dramatically~\cite{ongaro2014search, duan2018beat}. The difficulty stems from the greatly varying assumptions used in the various algorithms (e.g., failure and network assumptions), the inconsistent criteria of complexity metrics (e.g., message and time complexities), and the lack of comprehensive, standard, and intuitive protocol descriptions (e.g., message-passing patterns and workflows).

In this paper, we raise these challenges by providing comprehensive, step-by-step, and intuitive descriptions of each surveyed BFT approach. We aim to significantly assist readers, including researchers and practitioners, in understanding, analyzing, and evaluating the various state-of-the-art BFT consensus algorithms. Our systematic method follows a structured approach to break down intricate BFT approaches into basic components. Each survey consists of the following components:

First, we present the \textit{system model and service properties} of each BFT approach, covering aspects such as the failure model, network model, safety, and liveness properties.

We then highlight unique features, considering how efficiency, robustness, or availability is addressed. The consensus workflow is described based on whether an algorithm utilizes a leader node, a leader-based or leaderless structure.

\begin{itemize}
    \item \textit{Leader-based BFT:} If the surveyed paper involves a leader, our survey provides two subsections introducing two types of consensus:
    \begin{itemize}
        \item \textit{Replication consensus:} We show how a transaction is committed, illustrating the lifecycle of a transaction from being proposed by a client to being committed by the system.

        \item \textit{Leadership consensus:} We explain how a leader is determined, addressing view changes or leader elections.
    \end{itemize}

    \item \textit{Leaderless BFT:} If the surveyed paper does not involve a leader, our survey offers two subsections detailing:
    \begin{itemize}
        \item \textit{Replication consensus:} We illustrate how a transaction is committed without the coordination of a leader.
        
        \item \textit{Conflict resolution:} We show how conflicts are resolved in leaderless approaches without a global leader.
    \end{itemize}
\end{itemize}

Finally, we conclude each surveyed approach with a list of its \textit{pros and cons}. We also compare the surveyed approach with other similar methods (if any), addressing their respective strengths and weaknesses. In addition, each surveyed protocol is provided with visual summaries that depict the workflow of the consensus process. These visual representations significantly enhance the understanding of the original approaches, making them more accessible and facilitating the implementation process.

Compared to previous work that has surveyed BFT algorithms, such as~\cite{cachin2017blockchain, bach2018comparative, bano2019sok, distler2021byzantine}, we delve into the core BFT algorithms that achieve consensus and fault tolerance. Cachin et al.~\cite{cachin2017blockchain} present blockchain platforms and BFT consensus protocols proposed before 2015. Bano et al.~\cite{bano2019sok} survey Proof-of-X protocols and hybrid protocols used in permissionless blockchains. Distler~\cite{distler2021byzantine} provides a broad view of BFT-related topics, including the overall architectures of BFT solutions, agreement stages, checkpointing, failure recovery, and trusted subsystems. In contrast, our paper is dedicated to tackling the core consensus problem, demystifying intricate BFT mechanisms, and improving the understandability of state-of-the-art BFT algorithms. 

The remainder of this paper is organized as follows:
Section~\ref{sec:background-and-system-models} presents the background of commonly used assumptions of BFT consensus;
Section~\ref{sec:efficiency} introduces more efficient BFT algorithms that prioritize efficiency in replication and achieve high performance;
Section~\ref{sec:robustness} surveys more robust BFT algorithms that fortify robustness and defend against various attacks;
Section~\ref{sec:availability} presents more available BFT algorithms that can operate under asynchrony; and Section~\ref{sec:future} discusses future research directions in BFT consensus.

\section{Background}
\label{sec:background-and-system-models}

This section serves as an introduction to the fundamental concepts underpinning the topic of BFT consensus. BFT consensus algorithms play a crucial role in distributed systems by ensuring agreement among nodes in the presence of maliciously faulty components. In this overview, we delve into several key elements of BFT consensus algorithms, shedding light on the commonly used system models and underlying assumptions.

\textbf{The consensus problem and state machine replication.}
The consensus discussion often relates to \emph{state machine replication} (SMR) in distributed systems~\cite{lynch1981describing, owicki1976axiomatic, owicki1982proving, schneider1990implementing}. State machines serve as models for a collection of server replicas functioning coherently to provide one or more services to clients. A state machine typically comprises two fundamental components: \emph{state variables} and \emph{commands}. The state variables represent the current states of the system, while the commands represent the transitions between different states~\cite{schneider1990implementing}. 

Consensus algorithms synchronize and coordinate multiple state machines to ensure each machine computes the same result. As such, these machines effectively function as a logical single server, providing a unified and consistent view of system states. Consensus algorithms maintain this property even when some machines experience failures.

\begin{table}[t!]
    \renewcommand{\arraystretch}{1.1}
    \begin{adjustbox}{max width=\linewidth}
  \begin{tabular}{|c|c|c|c|c|}
    \hline

    & Algorithm & Msg Cplx. ($|M|$ entries) & Msg Cplx. ($f$ views) & Time Cplx.\\
    \hline
    \multirow{8}{*}{\rotatebox[origin=c]{90}{\parbox[c]{2.5cm}{\centering \small More Efficient}}} & PBFT~\cite{castro1999practical}/BFT-SMaRt~\cite{bessani2014state} & $O(|M|n^2)$ & $O(fn^3)$ & $5$ \\
    & Zyzzyva~\cite{kotla2007zyzzyva}$^{1}$ & $O(|M|n)$ & $O(fn^3)$ & $5$ \\
    & SBFT~\cite{gueta2019sbft}       & $O(|M|n)$   & $O(fn^2)$ & $6$ \\
    & HotStuff/LibraBFT~\cite{yin2019hotstuff} & $O(|M|n)$ & $O(fn)$ & $8$ \\
    & DAG-Rider~\cite{keidar2021all}$^{3}$   & $O(|M|n^3 \log n)$  & $O(fn)$ & $5+$ \\
    & Narwhal~\cite{danezis2022narwhal}$^{3}$   & $O(|M|n^3 \log n)$  & $O(fn)$ & $5+$ \\

    \cline{1-5}
    \multirow{5}{*}{\rotatebox[origin=c]{90}{\parbox[c]{2cm}{\centering \small More Robust}}} & Aardvark~\cite{clement2009making}$^{4}$ & $O(|M|n^2)$ & $O(fkn^3)$ & $5$\\
    & Pompe~\cite{zhang2020byzantine}$^{5}$ & $O(|M|n)+X$  & $X$ & $4+X$ \\
    & Spin~\cite{spin2009} & $O(|M|n^2)$ & $O(fn^2)$ & $7$ \\
    & RBFT~\cite{aublin2013rbft} & $O(|M|n^3)$ & $O(fn^4)$ & 7\\
    & Prosecutor~\cite{zhang2021prosecutor} & $O(|M|n)$ & $O(fn^2)$ & $5$ \\
    & PrestigeBFT~\cite{zhang2023prestigebft} & $O(|M|n)$ & $O(fn^2)$ & $7$ \\

    \hline
    \multirow{5}{*}{\rotatebox[origin=c]{90}{\parbox[c]{2.5cm}{\centering More Available}}} &  DBFT~\cite{crain2018dbft}$^*$ & $O(|M|n^2 + |c|n^3)$ & N/A & $\log n$ \\
    & HoneyBadgerBFT~\cite{miller2016honey}$^*$ & $O(|M| n + |c|n^3\log n)$ & N/A & $ \log n$ \\
    & BEAT0~\cite{duan2018beat}$^*$ & $O(|M| n + |c|n^3\log n)$ & N/A & $\log n$ \\
    & BEAT1~\cite{duan2018beat}$^*$ and BEAT2~\cite{duan2018beat}$^*$ & $O(|M| n^2 + |c|n^3\log n)$ & N/A & $\log n$ \\
    & BEAT3~\cite{duan2018beat}$^*$ and BEAT4~\cite{duan2018beat}$^*$ & $O(|M| + |c|n^3\log n)$ & N/A & $\log n$ \\
    \bottomrule
  \end{tabular}
  \end{adjustbox}
\begin{flushleft}
{
\footnotesize
$1$ Zyzzyva obtains an $O(n)$ message complexity only in the optimal path; otherwise, the message complexity is $O(n^2)$.\\ 
$2$ DAG-Rider and Narwhal use DAG-based mechanisms to achieve consensus; the minimum time complexity is $5$ rounds. \\
$3$ Aardvark invokes a view change if a leader fails to meet expected throughput; it may undergo more than $f$ view changes.\\
$4$ Pompe achieves full consensus by relying on its underlying consensus algorithm, the complexity of which is denoted by $X$.\\
$^*$ DBFT, HoneyBadgerBFT, and BEATs use the binary Byzantine protocol of a $|c|$ message complexity.
}
\end{flushleft}

\caption{Qualitative comparisons among state-of-the-art BFT algorithms in terms of message complexity in normal operation and view changes as well as time complexity.}
  \label{tab:comparison-all-bft}
  \vspace{-2em}
\end{table}

\textbf{Byzantine failures.} Byzantine failures are the strongest form of failures in the context of SMR. Under the BFT model, servers can exhibit arbitrary behavior, including changing message content and mutating their states~\cite{attiya2004distributed}. A faulty server may intentionally collude with the other faulty servers and jointly exhibit malicious behavior~\cite{lamport2019byzantine}. Byzantine failures differ from \emph{benign failures}~\cite{schneider1984byzantine}, where the message content is immutable and the change of the state of faulty servers is detectable.

\textbf{Byzantine quorums}. The decision-making in BFT algorithms often relies on quorum certificates~\cite{gifford1979weighted}, such as replicating a transaction or electing a leader server. A quorum certificate is a set of identical messages collected from different participants. The minimum quorum sizes to tolerate benign failures (e.g., Paxos~\cite{lamport1998part}, Raft~\cite{ongaro2014search}, and Escape~\cite{zhang2022escape}) and Byzantine failures (e.g., PBFT~\cite{castro1999practical} and HotStuff~\cite{yin2019hotstuff}) are $f+1$ and $2f+1$, respectively. Byzantine quorums introduce more redundancy to tolerate a stronger form of failure. Under BFT, since $f$ nodes are assumed faulty, the system must be flexible enough to exclude $f$ nodes in a quorum. Additionally, a quorum must at least have $f+1$ non-faulty nodes in case $f$ faulty nodes are included. Therefore, the minimum size of Byzantine quorums is $2f+1$ in a total of $n=3f+1$ nodes.

\textbf{Network assumption.} Consensus algorithms may operate under different network assumptions, which are often classified as \emph{synchronous}, \emph{asynchronous}, and \emph{partially synchronous}~\cite{dwork1988consensus}. Synchronous networks have a fixed upper bound (denoted by $\Delta$) on the time for message delivery and a fixed upper bound (denoted by $\delta$) of the discrepancy of processors' clocks, which allows executions to be partitioned into rounds. In contrast, asynchronous networks have no fixed upper bound of message delivery (i.e., $\Delta$ does not exist) or the discrepancy of processors' clocks (i.e., $\delta$ does not exist)~\cite{fischer1985impossibility}. 
In between the two assumptions, communication among servers can have a \emph{global stabilization time} (GST), unknown to processors. A network is partially synchronous if $\Delta$ and $\delta$ both exist but are unknown, or $\Delta$ and $\delta$ are known after the GST~\cite{dolev1987minimal, dwork1988consensus}.

\textbf{Complexity measures.} Two complexity measurements are commonly used to evaluate the efficiency of consensus algorithms: message complexity and time complexity. The \emph{message complexity} of a consensus algorithm is the total number of messages sent. For example, in an $n$-server cluster whose communication topology is a complete (fully connected) graph, if a server broadcasts a message for $k$ rounds, the message complexity is $O(kn)$. If every server broadcasts a message for $k$ rounds, the message complexity is $O(kn^2)$.

In addition, \emph{time complexity} measures the number of message-passing rounds that an execution of consensus takes. For example, if server $S_i$ sends a message to server $S_j$ and then $S_j$ replies to $S_i$, the time complexity is $2$. When the network is asynchronous, measuring the time complexity requires a finite delay time for message transmission~\cite{peterson1977economical}.

\textbf{Service properties.} Consensus algorithms coordinate server actions and provide correctness guarantees. In particular, an $f$-resilient BFT consensus algorithm needs to meet three criteria: \emph{termination}, \emph{agreement}, and \emph{validity}~\cite{pease1980reaching, lamport2019byzantine}.

\begin{description}
    \item[Termination.] Every non-faulty server eventually decides on a value.

    \item[Agreement.] Non-faulty servers do not decide on conflicting
    values; i.e., no two non-faulty servers decide differently.
    
    \item[Validity.] If all servers have the same input, then any
    value decided by a non-faulty server must be that common input.
\end{description}

The correctness guarantee can also be interpreted by \emph{safety}
and \emph{liveness}~\cite{robert1967assigning, lamport1977proving}. 

\begin{description}
    \item[Safety.] No two non-faulty replicas agree differently on a total order for the execution of requests despite failures.
    \item[Liveness.] An execution will terminate if its input is correct.
\end{description}

The safety property ensures that something will not happen: Correct replicas agree on the same total order for the execution of requests in the presence of failures. The liveness property, however, guarantees that something must happen: An execution eventually terminates if its input is correct; in the context of SMR, clients eventually receive replies to their requests.

\section{Toward more efficient BFT consensus}
\label{sec:efficiency}

Efficiency of replication is the highest priority for most BFT applications, including permissioned blockchain platforms such as
HyperLedger~\cite{androulaki2018hyperledger}, CCF~\cite{shamis2022ia}, and Diem~\cite{diem}. After PBFT~\cite{castro1999practical} pioneered a practical solution for BFT consensus, numerous approaches have proposed extensions and optimizations to improve system performance~\cite{castro1999practical, gupta13resilientdb, kotla2007zyzzyva, kotla2004high, li2007beyond, guerraoui2010next,yin2019hotstuff, zhang2021prosecutor, zhang2020byzantine,buchman2016tendermint}. Among these optimizations, leader-based BFT algorithms have gained favor due to their efficient coordination in the consensus processes during normal operation. In this section, we first present a detailed survey of PBFT~\cite{castro1999practical} in \S\ref{sec:pbft}, the foundation of modern BFT consensus algorithms. Then, as shown in Figure~\ref{fig:taxonomyofbft}, the rest of this section consists of BFT protocols that reduce message complexity, such as SBFT~\cite{gueta2019sbft} and HotStuff~\cite{yin2019hotstuff} (in \S\ref{sec:sbft} and \S\ref{sec:hotstuff}), utilize multi-leader consensus, such as ISS~\cite{stathakopoulou2022state} (in~\S\ref{sec:iss}), and enable DAG-based pipelines, such as DAG-Rider~\cite{keidar2021all} and Narwhal and Tusk~\cite{danezis2022narwhal} (in \S\ref{sec:dagrider} and \S\ref{sec:narwhal}). In this section, we present comprehensive, step-by-step descriptions of these algorithms and conduct a thorough analysis of the features that render them more efficient. 

\subsection{PBFT: Practical Byzantine fault tolerance}
\label{sec:pbft}
\subsubsection{System model and service properties}
PBFT pioneered a practical solution that achieves consensus in the presence of Byzantine faults. PBFT guarantees safety when there are at most $f$ faulty replicas out of a total of $n=3f+1$ replicas; the safety guarantee does not rely on any assumption of network synchrony; i.e., safety is provided in asynchronous networks. However, PBFT requires synchrony with bounded message delays to guarantee liveness. The replication protocol operates through a progression of views numbered with consecutive integers. The view-change protocol first selects a replica as a primary in a view, and the other replicas assume a backup role. The parameters of all messages involved in the replication and view-change protocols are presented in Table~\ref{tab:PBFT_messaging_formats}.

\begin{figure*}[t]
    \begin{subfigure}{0.49\textwidth}
        \centering
        \includegraphics[width=0.9\linewidth]{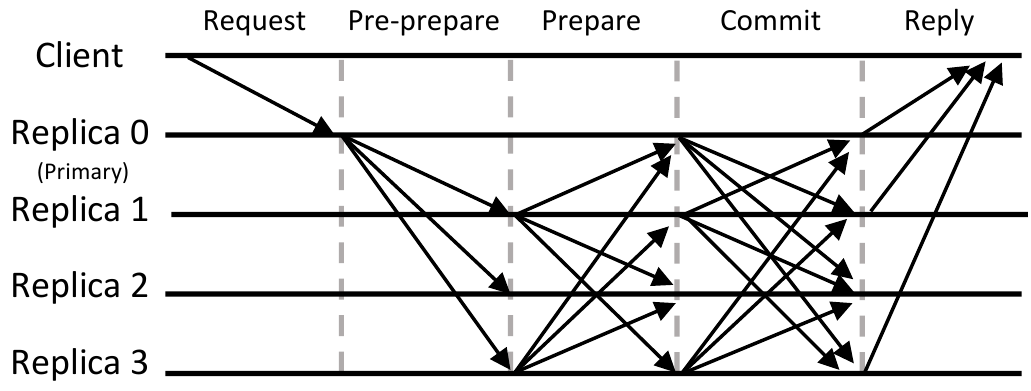}
        \caption{The replication protocol under normal operation.}
        \label{fig:pbft-normal-operation}
    \end{subfigure}
    \hfill
    \begin{subfigure}{0.49\textwidth}
        \centering
        \includegraphics[width=0.9\linewidth]{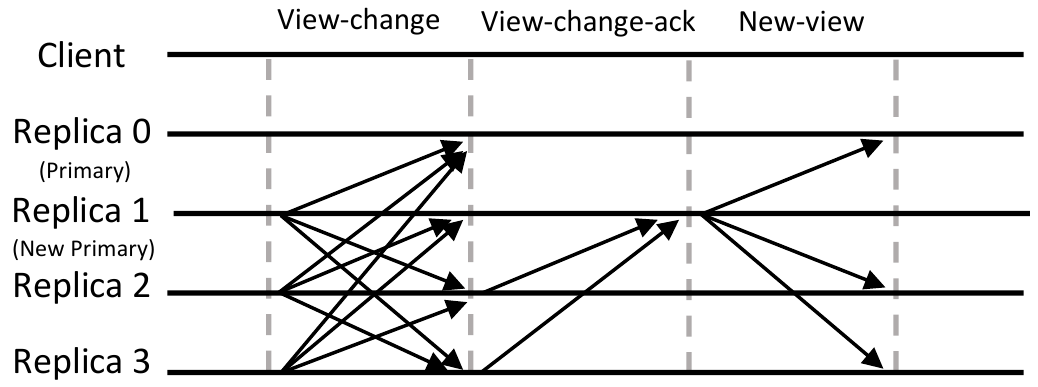}
        \caption{The view-change protocol under primary
          failure.}
        \label{fig:pbft-view-change}
    \end{subfigure}
    \caption{The message-passing workflow of normal operation and view changes in a four-replica PBFT system.}
    \label{fig:pbft}
\end{figure*}

\subsubsection{The replication protocol}
The consensus in replication consists of an array of linearizable consensus instances for each client request. In each consensus instance, replicas make \emph{predicates} for the primary's instructions in each phase based on quorums that include votes from $2f+1$ different replicas. The replication also uses \emph{checkpoints} to periodically confirm that the requests committed by terminated consensus instances have been successfully executed.

Specifically, a consensus stance starts when a client invokes an operation to a primary. As illustrated in Figure~\ref{fig:pbft-normal-operation}, a client requests to invoke an operation (\texttt{op}) on the primary, which starts the first phase in replication, \emph{the request phase}. This operation invokes a PBFT service, such as an invocation to propose or query a value (i.e., a write or read), and each operation has a unique timestamp. 
After sending a \textsc{request} message, the
client starts a timer, waiting for the completion of the requesting
operation: if the client receives $f+1$ \textsc{reply} messages from
different replicas, the client considers the operation to be completed
and stops the timer. However, if the client cannot receive a
sufficient number of replies before the timer expires, this implies
that the invocation of the requesting operation may have failed. This
failure can be caused by \textit{internal faults} (e.g., the
primary failed) and \textit{external faults} (e.g., the request failed
to be delivered to the primary). Clients cannot deal with the
former faults and must simply wait for the system to
recover. However, the latter faults can be handled by a different
messaging scheme; if a client's timer expires, the client resends the
request message by broadcasting it to all replicas. Replicas that
receive such a request transmit it to the primary. Although this
scheme increases the messaging complexity of the request phase from
$O(1)$ to $O(n)$, it allows the client to invoke operations when the
link between the client and primary is not reliable.

After receiving a request, the primary starts to conduct consensus among replicas to commit the request. The consensus starts with \emph{the pre-prepare phase} in which the primary assigns a unique sequence number to the request. The sequence number is chosen from the range of $h$ to $h+k$, where $h$ is the sequence number of the last stable checkpoint, and $k$ is a predefined value used to limit the growth of sequence numbers. The limited growth prevents a faulty primary from exhausting the space of the sequence numbers. Then, the primary assembles the current view number, sequence number, and digest of the request message into a \textsc{pre-prepare} message and then signs it; the \textsc{pre-prepare} message also piggybacks the original request received from the client, and the primary sends the message to all backups.

Next, the process enters \emph{the prepare phase}. After receiving the \textsc{pre-prepare} message, backups \textcircled{1} verify the signature, \textcircled{2} check that they are in the same view as the primary, \textcircled{3} confirm that the sequence number has not been assigned to other requests, and \textcircled{4} compute the digest of the piggybacked request to ensure the digests match. If the message is valid, backups proceed by broadcasting \textsc{prepare} messages to all replicas. Then, a backup collects \textsc{prepare} messages from the other backups. If the backup receives $2f$ \textsc{prepare} messages from different backups, the backup makes a predicate that the request is \emph{prepared}. The predicate is supported by a \emph{quorum certificate} consisting
of $2f+1$ agreements (including itself) of the same order from
different replicas. Since there are at most $f$ faulty replicas among
the total of $n=3f+1$ replicas, the combination of pre-prepare and
prepare phases ensures that non-faulty replicas have reached an
agreement on the order of the request in the same view.

\begin{table}[t]
    \centering
    \small
    \begin{tabular}{|r|l|}
        \hline
        Message types & Parameters piggybacked in messages\\
        \hline
        \textsc{request} & $\langle$
        \texttt{op,timestamp,clientId}
        $\rangle_{\sigma_c}$\\
        
        \textsc{pre-prepare} & $\langle \langle$
        \texttt{view,seqNumber,digestOfRequest}
        $\rangle_{\sigma_P}$, \texttt{request}$\rangle$ \\
        
        \textsc{prepare} & $\langle$
        \texttt{view,seqNumber,digestOfRequest,serverId}
        $\rangle_{\sigma_i}$\\
        
        \textsc{commit} & $\langle$
        \texttt{view,seqNumber,digestOfRequest,serverId}
        $\rangle_{\sigma_i}$\\
        
        \textsc{reply} & $\langle$
        \texttt{view,timestamp,clientId,serverId,resultOfOp}
        $\rangle_{\sigma_i}$\\
        \hline
        
        \textsc{checkpoint} & $\langle$ 
        \texttt{seqNumber,digestOfState,serverId}
        $\rangle_{\sigma_i}$\\
        \hline
        
        \textsc{view-change} & $\langle$ 
        \texttt{newView,seqNumber,$\mathcal{C}$,$\mathcal{P}$,$\mathcal{Q}$,serverId}
        $\rangle_{\sigma_i}$\\
        \textsc{view-change-ack} & $\langle$
        \texttt{newView,serverId,vcSenderId,digestOfVcMsg}
        $\rangle_{\sigma_i}$\\
        \textsc{new-view} & $\langle$
        \texttt{newView,$\mathcal{V}$,$\mathcal{X}$}
        $\rangle_{\sigma_P}$\\
        \hline
    \end{tabular}
    \caption{The messaging formats in PBFT.}
    \label{tab:PBFT_messaging_formats}
    \vspace{-2em}
\end{table}

After the prepared predicate is made, the replica informs the other
replicas about the predicate in \emph{the commit phase} by
broadcasting a \textsc{commit} message. Similar to the previous phase,
replicas make a predicate of the request as \emph{committed} when $2f$
\textsc{commit} messages have been received from different replicas;
the redundancy forms a quorum certificate that $2f+1$ replicas have
committed the same request in the agreed order. The quorum certificate
also guarantees a \emph{weak certificate}, which supports a
\emph{committed-local} predicate that $f+1$ replicas have committed
the request. By the end of the commit phase, the quorum certificate
provides the safety property: no two correct replicas execute the
request differently.

Finally, when predicate \emph{committed} is valid, a backup starts
\emph{the reply phase} and sends the client a \textsc{reply} message
with a Boolean variable (\texttt{resultOfOp}=true), indicating that the
consensus for committing the proposed request is reached. The client
waits for a weak certificate to confirm the result; that is, it waits
for $f+1$ replies with valid signatures from different replicas.

Once a request is committed, the execution of the request reflects the state of a replica. The replica must show the correctness of its states when sharing them with other replicas. Instead of generating proofs for every execution, the algorithm periodically computes proofs for an array of executions, namely \emph{checkpoints}, as a way to show states. To produce checkpoints, a replica broadcasts a \textsc{checkpoint} message. If $2f+1$ \textsc{checkpoint} messages are collected, the proof of the checkpoint is formed, and the checkpoint becomes a \emph{stable checkpoint}. The sequence number of the stable checkpoint is denoted by $h$ as a low water mark. A high water mark is $H=h+k$ where $k$ is a sufficiently large predefined number. The low and high water marks bound the growth of sequence numbers, as introduced in the pre-prepare phase. After obtaining a stable checkpoint, the intermediate log entries for creating previous quorum certificates can be discarded. The incremental computation of checkpoints can clean log entries appended at a previous checkpoint. Thus, this progressive oblivation of stale log entries achieves garbage collection and reduces space overhead.


\subsubsection{The view-change protocol}
When a primary fails, the coordination of consensus is impeded, which puts liveness in jeopardy. PBFT handles primary failures by its view-change protocol, and the ability to progress in the succession of views underpins liveness. Specifically, if the primary of view $v$ fails, the algorithm proceeds to view $v+1$ and selects a new primary such that ${P=(v+1) \mod n}$. During a view change, replicas may have various replication states because replication works in an asynchronous premise; i.e., replicas may not be fully synchronized. Thus, this variety produces many tricky corner cases that are difficult to handle, bringing challenges to design and implementation. PBFT has different view-change protocols presented in the literature~\cite{castro1999practical, castro2001practical.thesis, castro2002practical}. We summarize the version in~\cite{castro2002practical}, which improves the version presented in~\cite{castro1999practical} but may require unbounded spaces. 

A view-change process has three phases (shown in Figure~\ref{fig:pbft-view-change}). In view $v$, initially, a backup starts a timer after receiving a valid request if the timer is not running at the moment. Then, the backup starts to wait for the request to be committed. The request can be received from the primary or clients. The backup stops its timer when the request is committed but restarts the timer if there are outstanding valid requests that have not been committed. If its timer expires, the backup considers the primary failed and prepares to enter view $v+1$ by starting the \emph{view-change phase}. It broadcasts a \textsc{view-change} message, where $\texttt{newView}=v+1$ and the sequence number is the latest stable checkpoint; $\mathcal{C}, \mathcal{P}$, and $\mathcal{Q}$ are indicators of the replication progress in view $v$: $\mathcal{C}$ contains the sequence number and digests of each checkpoint, and $\mathcal{P}$ and $\mathcal{Q}$ store information about requests that have prepared and pre-prepared between the low and high water marks, respectively. 

After receiving a \textsc{view-change} message, a replica verifies the message such that $\mathcal{P}$ and $\mathcal{Q}$ are produced in views that are less than $v$. Then, the replica sends a \textsc{view-change-ack} message to the primary of view $v+1$. The \textsc{ack} message contains the current replica ID, the digest and the sender of the received \textsc{view-change} message. The primary of view $v+1$ keeps collecting \textsc{view-change} and \textsc{view-change-ack} messages. After receiving $2f-1$ \textsc{view-change-ack} messages that acknowledge a backup $i$'s \textsc{view-change} message, the primary adds this \textsc{view-change} message to a set $\mathcal{S}$ and thus a quorum certificate, called \emph{view-change certificate}, is formed. 

Next, the new primary pairs the replica ID with its \textsc{view-change} message in $\mathcal{S}$ to a set $\mathcal{V}$ and selects the highest checkpoint, $h$, that is included in at least $f+1$ messages in $\mathcal{S}$. The primary extracts requests whose sequence numbers are between $h$ and $h+k$ and have been prepared in view $v$ or pre-prepared by a quorum certificate and adds these requests into a set $\mathcal{X}$. Sets $\mathcal{V}$ and $\mathcal{X}$ allow requests that are uncommitted but seen by a quorum in the previous view to be processed in the new view.

Finally, backups check whether the messages in $\mathcal{V}$ and $\mathcal{X}$ support the new primary's decision to choose checkpoints and the continuance of uncommitted requests. If not, the replicas start a new view, moving to view $v+2$. Otherwise, the legitimacy of the new primary is confirmed while the view-change protocol terminates and normal operation resumes.

\subsubsection{Pros and cons}
PBFT pioneered a practical solution for Byzantine fault tolerance in asynchronous networks and made a significant impact as the beginning of an era of efficient BFT algorithms. The extended version of PBFT~\cite{castro2002practical} also provided a proactive recovery approach to recover faulty replicas. PBFT was adopted by many permissioned blockchains, although it was not designed to cope with many failures. Numerous works are inspired by PBFT to obtain higher throughput and lower latency consensus (e.g., BFT-SMaRt~\cite{bessani2014state}, SBFT~\cite{gueta2019sbft}, HotStuff~\cite{yin2019hotstuff}, and Prosecutor~\cite{zhang2021prosecutor}). 

However, PBFT's quadratic messaging complexity hinders it from use in applications at large scales~\cite{brewer2000towards, thakkar2018performance}. Furthermore, the client interaction may lead the system to repeated view changes without making progress if faulty clients use an inconsistent authenticator on requests~\cite{clement2009making}.

\subsection{SBFT: A scalable and decentralized trust infrastructure}
\label{sec:sbft}
 \subsubsection{System model and service properties}
 SBFT~\cite{gueta2019sbft} is a leader-based BFT consensus algorithm that tolerates $f$ Byzantine servers and $c$ crashed or straggler servers by using $3f+2c+1$ servers in total. Similar to Zyzzyva~\cite{kotla2007zyzzyva}, SBFT enables a \emph{fast path} for achieving consensus if no failure occurs and the network is synchronous. Otherwise, SBFT can seamlessly switch to a slow path, namely \emph{linear-PBFT}, featuring a dual model without engaging view changes. Compared with PBFT~\cite{castro1999practical}, SBFT avoids the quadratic Byzantine broadcast (i.e., $n$-to-$n$ messaging) by utilizing threshold signatures; threshold signatures convert an array of signed messages into one threshold signed message where a threshold indicates the number of signers. By utilizing threshold signatures, under normal operation, SBFT achieves linear consensus with a message transmission cost of $O(n)$.
 
\begin{figure*}[t]
    \begin{subfigure}{0.49\textwidth}
        \centering
        \includegraphics[width=\linewidth]{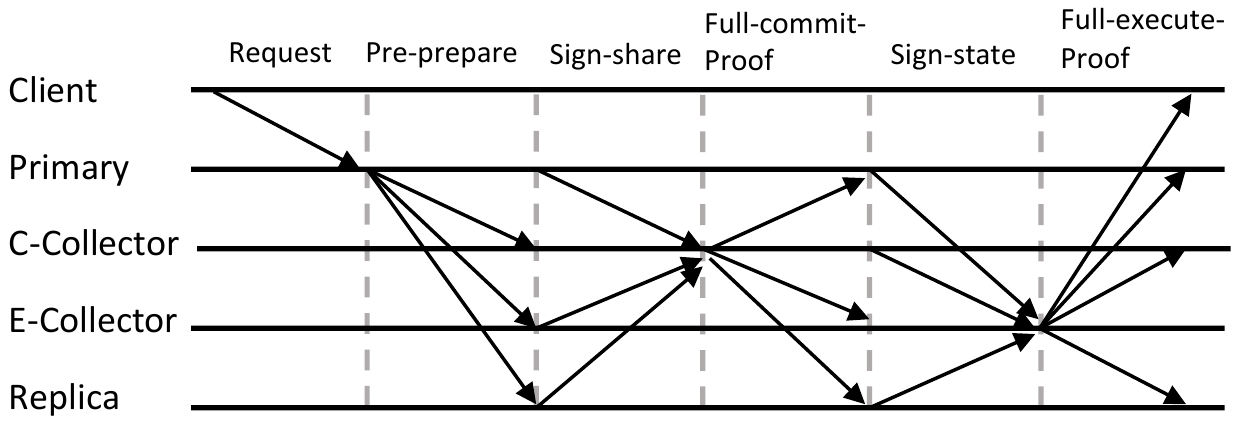}
        \caption{Replication in normal operation.}
        \label{fig:sbft-normal-operation}
    \end{subfigure}
    \hfill
    \begin{subfigure}{0.49\textwidth}
        \centering
        \includegraphics[width=\linewidth]{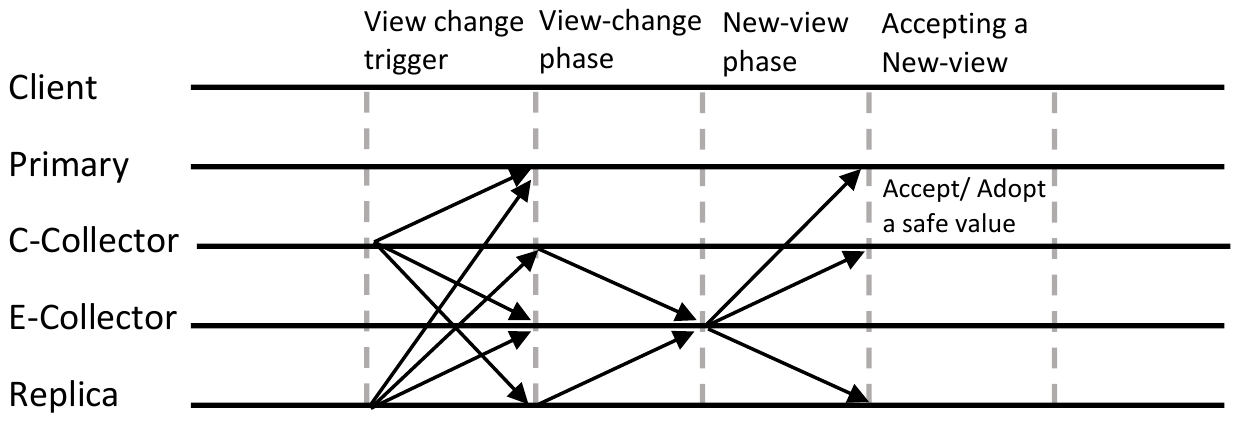}
        \caption{View changes under primary failures.}
        \label{fig:sbft-view-change}
    \end{subfigure}
    \caption{The message-passing in normal operation and view change in SBFT when $n=4$, $f=1$, $c=0$.}
\end{figure*}

\subsubsection{The replication protocol}
 SBFT uses different servers acting as ``local'' primaries in different phases, amortizing the coordination workload among a group of selected collectors.  In particular, as shown in Figure~\ref{fig:sbft-normal-operation}, SBFT delineates three key phases under normal operation; in each phase, messages are broadcast and collected by a single server. The fast path mode of SBFT, in which the network is synchronous and no servers fail, makes use of this messaging scheme to achieve fast consensus. 
 
 \textbf{Fast path.}
 First, in the \texttt{pre-prepare} phase, the primary manages the order of transactions proposed by clients and broadcasts instructions to all other servers.
 Then, after receiving the instruction from the primary, servers validate the primary's message and prepare a reply message; instead of replying to the primary, servers send the reply message to the commit collector (\emph{C-collector}). 
 Next, in the \texttt{full-commit-proof} phase, the C-collector takes the primary responsibility by collecting signed replies, converting them into one threshold signed message, and broadcasting the message to all other servers.
 Finally, in the \texttt{full-execute-proof} phase, the leadership duty is assigned to the execution collector (\emph{E-collector}).
 The E-collector collects replies, aggregates them into one threshold signed message, and disseminates them to the others including the proposing client. Compared with PBFT, whose clients confirm the commit of a proposed transaction based on $f+1$ messages, SBFT sends only one confirm message to a proposing client.
 
 \textbf{linear-PBFT.}
 In contrast to the fast path mode, under the presence of failures or partial synchrony, SBFT switches to the linear-PBFT mode. In this mode, regardless of $n$, there are at most $2$ collectors or all collector roles are aggregated to the primary. 
 The reason this model is called linear-PBFT is that if SBFT uses only the primary as all ``local'' leaders in each phase, the workflow becomes similar to PBFT~\cite{castro1999practical}. 
 Nonetheless, the messaging complexity is reduced from $O(n^2)$ to $O(n)$, as each server receives an $O(1)$ threshold signed message in the linear-PBFT mode. 
 
 In a given view, SBFT selects a primary based on the view number and chooses collectors based on the commit state index (sequence number) and the view number. 
 SBFT suggests randomly selecting a primary and collectors to increase robustness, though the design and implementation are unprovided. 
 In the linear-PBFT mode, the primary is always chosen as the last collector. The transition of server roles is initiated by view changes, which select a new set of a primary and collectors when the current primary is faulty.
 
\subsubsection{The view-change protocol}
 SBFT's view change mechanism has a message transmission complexity of $O(n^2)$ (shown in Figure~\ref{fig:sbft-view-change}). A view change is initiated when a replica triggers a timeout or receives a proof that $f+1$ replicas suspect the primary is faulty. 
 This begins the \texttt{view-change trigger} phase; in the worst case, all servers trigger new views, resulting in an $O(n^2)$ message transmission.
 Then, in the \texttt{view-change} phase, each replica maintains the last stable sequence number, denoted by $ls$. 
 Since the network can be partially synchronous, concurrent in-process consensus may lead to discrepant $ls$ among servers; SBFT limits the number of outstanding blocks by a predefined parameter, denoted by $win$; thus, a sequence number $s$ lies between $ls$ and $ls + win$. 
 The sequence number indicates the state of a server; the servers that have triggered a view change send the new primary a view-change message consisting of $ls$ and digests of corresponding states from $ls$ to $ls + win$. 
 After the new primary solicits a set of $2f+2c+1$ view-change messages, denoted by $\mathcal{I}$, in the \texttt{new-view} phase, it initiates a new view and broadcasts $\mathcal{I}$ to all other servers.
 Then, in the \texttt{accepting a new-view} phase, servers accept the values indexed from slot $ls$ to $ls + win$ or adopt a \emph{safe value}, a symbol that the corresponding sequence number can be used for a future transaction as its index in the new view. 
 
\subsubsection{Pros and cons}
 By utilizing threshold signatures, SBFT achieves linear message transmission under normal operation in both fast path and linear-PBFT modes. Compared with PBFT, SBFT also reduces client communication from $O(f+1)$ to $O(1)$ since E-collectors aggregate server replies in the \texttt{full-execute-proof} phase. However, SBFT inherits some weaknesses from PBFT~\cite{clement2009making}. Faulty clients can trigger unnecessary view changes if they only partially communicate with $f+1$ replicas. In SBFT, a view change is triggered if $f+1$ replicas complain; faulty clients can exploit this feature by sending transactions to a set of $f+1$ replicas excluding the primary. After the $f+1$ replicas trigger timeout, a view change is initiated to replace the correct primary, invoking unnecessary leadership changes. 
\subsection{HotStuff: BFT consensus in the lens of blockchain}
\label{sec:hotstuff}
\begin{figure}[t]
    \centering
    \includegraphics[width=0.85\linewidth]{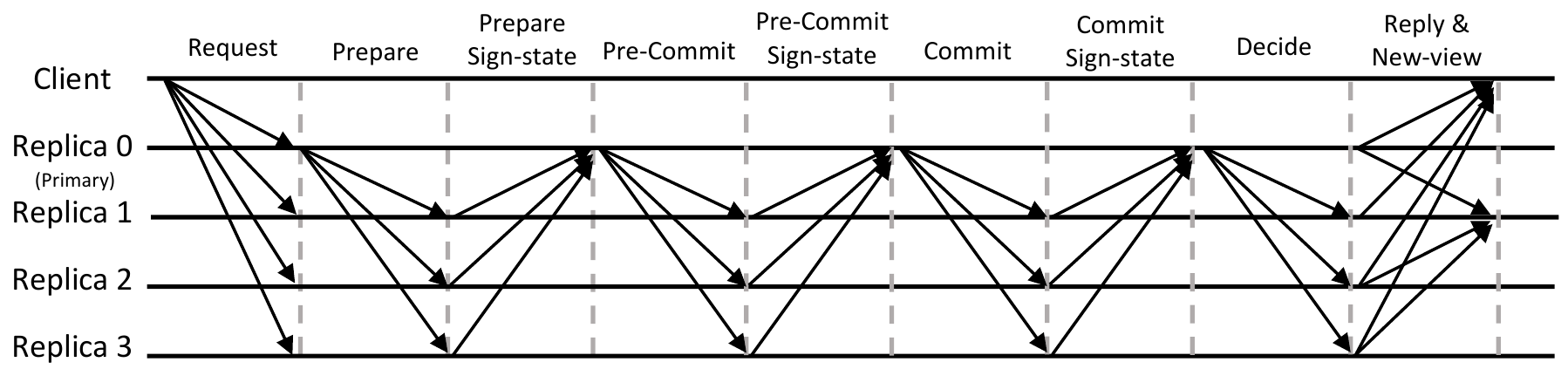}
    \caption{The workflow of the $4$-phase replication under normal operation in HotStuff.}
    \label{fig:hotstuff-normal-operation}
    \label{fig:hotstuff}
\end{figure}

 \subsubsection{System model and service properties}
 HotStuff~\cite{yin2019hotstuff} is a leader-based BFT consensus algorithm that tolerates up to $f$ Byzantine servers using a total of $3f+1$ servers. It assumes a partially synchronous network to achieve liveness, whereas safety does not rely on any assumption of network synchrony. HotStuff guarantees optimistic responsiveness; i.e., a consensus decision requires only the first $2f+1$ messages to make progress. In addition, HotStuff rotates primary servers for each consensus instance to ensure chain quality~\cite{garay2015bitcoin}. Similar to SBFT~\cite{gueta2019sbft}, HotStuff uses threshold signatures to obtain consensus linearity (i.e., $O(n)$ messaging complexity) under normal operation, reducing the communication overhead of Byzantine broadcasts that reach consensus by $n$-to-$n$ messaging in PBFT~\cite{castro1999practical}. However, in the worst case of cascading failed primaries, HotStuff takes up to $O(f*n)$ rounds to reach consensus, on the order of $O(n^2)$. 
 


\subsubsection{The replication protocol}
A replication process starts when a client broadcasts a request to invoke an operation to all replicas (shown in Figure~\ref{fig:hotstuff}). Similar to PBFT~\cite{castro1999practical}, the client waits for replies from $f+1$ replicas to confirm that an invoked operation is executed. HotStuff does not further discuss the handling of client failures but references standard literature for handling numbering and deduplication of client requests~\cite{castro1999practical, bessani2014state}.

HotStuff operates in the succession of views. Under normal operation, a consensus process has four phases (shown in Figure~\ref{fig:hotstuff}). After receiving a client request, the primary broadcasts a \textsc{prepare} message to all replicas. A replica verifies the message based on two criteria: \textcircled{1} the message extends from a previously committed message (they are continuous with no other messages in between), and \textcircled{2} it has a higher view number (\textit{viewNum}) than the current view number. If the message is verified, the replica signs a \textit{prepare-vote} message by its partial signature and sends the vote back to the primary. The primary waits for $2f+1$ votes to form a quorum certificate (\textit{QC}) for the \textsc{prepare} phase, denoted by \textit{prepareQC}. 
After a \textit{prepareQC} is formed, the primary starts a \textsc{pre-commit} phase by broadcasting a message with the \textit{prepareQC}. 
All replicas store a received \textit{prepareQC} in this phase and send a \textit{commit-vote} message signed by their partial signatures back to the primary. When the primary receives $2f+1$ votes, it forms a \textit{pre-commitQC} for this \textsc{pre-commit} phase. Then, it starts a \textsc{commit} phase by broadcasting messages with \textit{pre-commitQC} to all replicas.
Replicas reply to the primary with a \textsc{commit} vote and update a local variable, denoted by \textit{lockedQC}, that keeps counting the number of received \textit{QC}s for a request. 
When the primary receives $2f+1$ \textsc{commit} votes, it broadcasts a \textsc{decide} message to all replicas. Replicas consider the consensus of the current client request to be reached and then execute the operation in the request. Replicas then increment the \textit{viewNum} and send a new \textsc{new-view} message to the new primary. Replicas also send a confirmation to the proposing client, indicating that the request is locally executed. The client waits for $f+1$ messages to confirm that its proposed request is executed. 

Throughout a consensus process, each replica stores three key variables to reach consensus for a client request: \textit{prepareQC}, the highest locked \textsc{prepare} message that a replica knows; \textit{lockedQC}, the highest locked \textsc{commit} message that a replica knows; and \textit{viewNum}, the current round the replica is in, updated with each \textsc{new-view}. 

Unlike PBFT~\cite{castro1999practical} and other leader-based protocols that maintain stable leadership~\cite{bessani2014state, gueta2019sbft, kotla2007zyzzyva, zhang2021prosecutor}, HotStuff rotates leadership for each consensus instance. Each consensus instance, identified by a monotonically increasing \textit{viewNum}, begins with a \textsc{new-view} message with \textit{viewNum} and references to the highest quorum certificate from the most recent \textsc{prepare} round, \textit{prepareQC}. The primary for the next round collects $2f+1$ \textsc{new-view} messages to identify the highest preceding view that it extends from, based on collected \textsc{prepareQC}s. If the new primary cannot collect a \textsc{new-view} message from $2f+1$ in time, the other replicas will trigger timeouts and initiate a view change that rotates the primary role to the next server.

\subsubsection{The view-change protocol}
 With the motivation of ensuring chain quality~\cite{garay2015bitcoin} for blockchain applications, HotStuff enables frequent view changes compared to state-of-the-art leader-based protocols~\cite{gueta2019sbft, castro1999practical, kotla2007zyzzyva, zhang2021prosecutor}. Each consensus instance starts with a new primary that conducts consensus by broadcasting coordination messages and collecting threshold signatures. As a result, the view change is engraved at the core of the HotStuff protocol. HotStuff adopts PBFT's leadership rotation scheme where a new primary $p$ in a view $v$ is decided such that $p = v \mod n$ where $n$ is the number of total servers. In addition, a broad range of strategies could be used for rotating primaries in each phase. For instance, LibraBFT~\cite{baudet2019state}, a variant of HotStuff in the Libra blockchain, uses round-robin leadership selection among all replicas to choose a primary. 

\subsubsection{Pros and cons}
 By using threshold signatures, HotStuff achieves linear message transmission for reaching consensus. In each phase, the primary collects votes and builds threshold signatures that can be pipelined; pipelining of decisions can further simplify the protocol for building chained HotStuff, similar to Casper~\cite{buterin2017casper}, and reduces the time complexity in reaching consensus. Furthermore, HotStuff can be extended to permissionless blockchains using delay towers (e.g., ~\cite{motepalli2022decentralizing}).
 
 However, HotStuff's frequent leadership rotation becomes problematic in the presence of failures. Since each server conducts a consensus instance for client requests, under $f$ failed servers, HotStuff's throughput drops significantly because the $f$ failed servers cannot achieve consensus when they are assigned with primary duties~\cite{zhang2021prosecutor}.
\subsection{ISS: State Machine Replication Scalability Made Simple}
\label{sec:iss}

\begin{figure*}[t]
    \begin{subfigure}{0.33\textwidth}
        \centering
        \includegraphics[width=\linewidth]{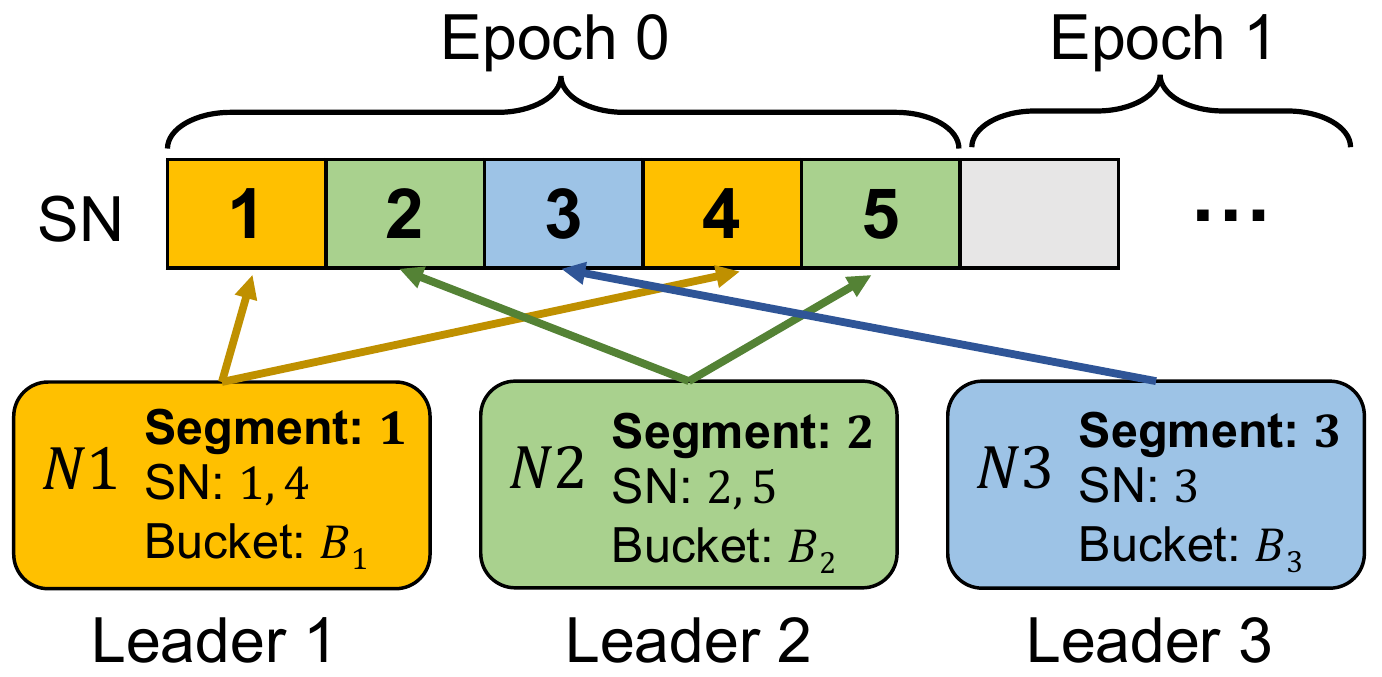}
        \caption{ISS partitions log into epochs. Each epoch is further partitioned into segments, with each segment consisting of unique sequence numbers and assigned to a leader node.}
        \label{fig:iss-segments}
    \end{subfigure}
    \hfill
    \begin{subfigure}{0.65\textwidth}
        \centering
        \includegraphics[width=\linewidth]{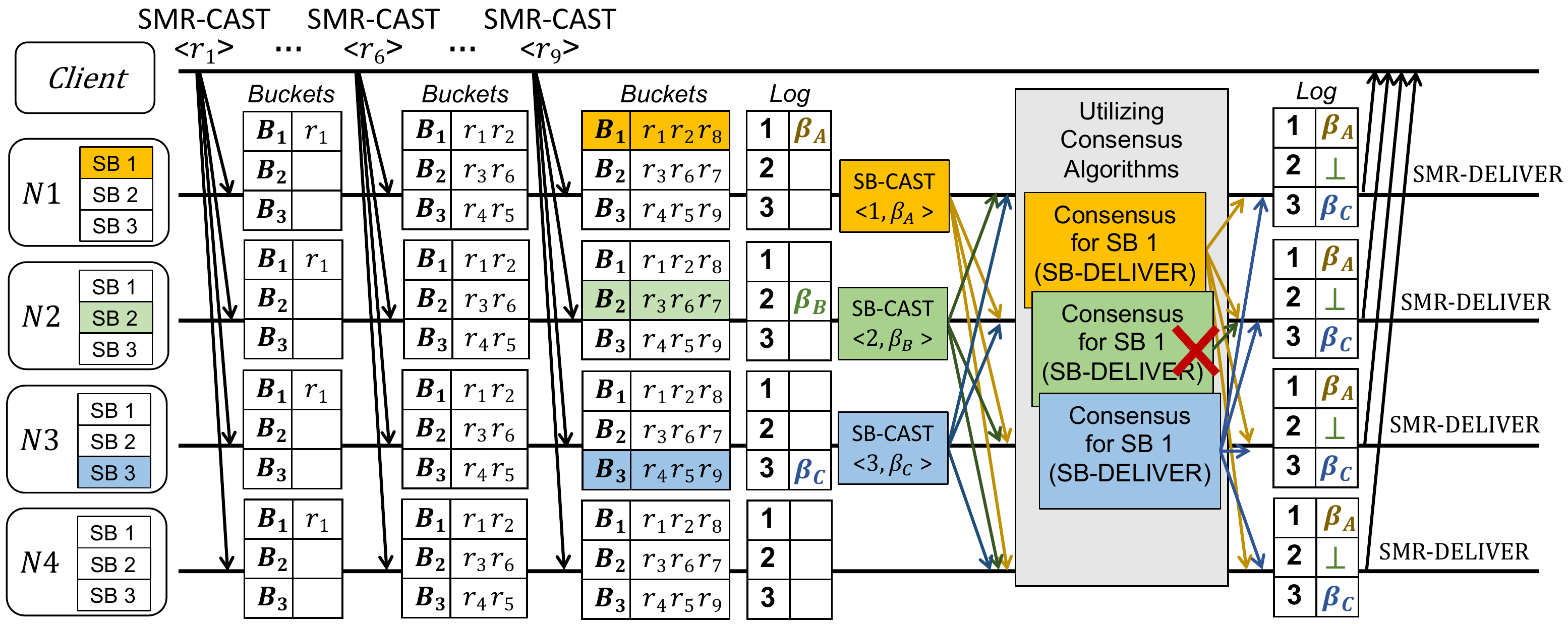}
        \caption{An example of parallel consensus for SB instances managed by multiple leaders, where $N2$ becomes faulty in the consensus phase.}
        \label{fig:iss-workflow}
    \end{subfigure}
    
    \caption{The workflow of ISS for utilizing multi-leader consensus.}
    \label{fig:pbft}
\end{figure*}

\subsubsection{System model and service properties}
Insanely Scalable State Machine Replication (ISS)~\cite{stathakopoulou2022state} introduces a generic framework that scales single-leader total order broadcast (TOB) protocols into a multi-leader protocol. This is achieved by partitioning the workload among multiple leaders and associating each with a Sequenced Broadcast (SB) instance that runs a single-leader TOB protocol; since SB instances can operate independently, this allows multiple TOB instances to run in parallel, thereby achieving a multi-leader protocol. ISS assumes a partially synchronous network and tolerates up to $f$ Byzantine faulty nodes among $3f+1$ nodes in total. The evaluation results show that ISS's multi-leader protocol outperforms single-leader protocols; its throughput is $37\times$, $56\times$, and $55\times$ higher than PBFT~\cite{castro1999practical}, Chained HotStuff~\cite{yin2019hotstuff}, and Raft~\cite{ongaro2014search}, respectively.

\subsubsection{The replication protocol}
In ISS, each node maintains a contiguous log to contain all accepted messages, as shown in Figure~\ref{fig:iss-segments}, and each message is comprised of client requests that are batched in order to decrease message complexity. Each position in the log corresponds to a sequence number (SN) based on its offset from the start of the log. The log is split into epochs containing a group of continuous sequence numbers, and all messages in an epoch must be processed before the next epoch may begin. In Figure~\ref{fig:iss-segments}, epoch $0$ consists of $5$ SNs: 1, 2, 3, 4 and 5. Each epoch is further partitioned into segments by assigning SNs to leader nodes in a round-robin fashion such that each segment is ultimately associated with one leader. Segments 1, 2 and 3 in Figure~\ref{fig:iss-segments} contain SN 1 and 4, 2 and 5, and 3, respectively. Further, all client requests received in ISS are sorted into buckets based on a modulo hash function. These buckets are assigned to leaders such that no leaders have the same bucket(s) and the same client request cannot be proposed multiple times.

ISS begins by selecting leader(s) based on a \textit{leader selection policy} (see~\S\ref{sec:iss:mls}) and gives each leader a segment in the current epoch. In Figure~\ref{fig:iss-segments}, the $5$ sequence numbers are divided into $3$ segments; nodes $N1$, $N2$, and $N3$ are selected as leaders and are associated with segments 1, 2, and 3, respectively. Each leader is allowed to make proposals by using only sequence numbers of its own segment. Thus, no two proposals can have the same sequence number. Next, each leader is assigned one bucket from which may take client request batches to use for proposals. In Figure~\ref{fig:iss-workflow}, $N1$, $N2$, and $N3$ are assigned buckets $B_{1}$, $B_{2}$, and $B_{3}$, respectively. 

Following the initial configuration, ISS begins to achieve consensus via multiple SB instances, each associated with one leader and its segment. In Figure~\ref{fig:iss-workflow}, $N1$, $N2$, and $N3$ are the leaders of concurrent SB instances 1, 2, and 3, respectively. Every node partakes in all SB instances but is the leader of at most one instance. Leaders receive client requests, denoted by <SMR-CAST|$r_{x}$>, and hash them into their local bucket queues. A leader waits until \textcircled{1} its bucket queue reaches a predefined batch size (e.g., the batch size in Figure~\ref{fig:iss-workflow} is $3$ (i.e., $|B_i|=3$)), or \textcircled{2} a predefined amount of time since the last proposal has elapsed. The leader then packs the queued requests into a batch (e.g., $\beta_A$) and broadcasts SB-CAST messages for its SB instance with one of its sequence numbers.

At this point, each leader runs the underlying wrapped TOB protocol (e.g., PBFT, HotStuff, or Raft) to achieve consensus for its proposed SB instance among all nodes. Since each SB instance is associated with different leader nodes, ISS empowers multiple SB instances to operate in parallel. For example, $N1$, $N2$, and $N3$ can conduct consensus for SB instances 1, 2, and 3 simultaneously in a non-blocking way. When consensus is reached for an SB instance, all non-faulty nodes sb-deliver the batch piggybacked in the SB instance. If consensus cannot be reached, a $\bot$ will be delivered to the log. For example, if $N2$ becomes faulty when it starts to sb-deliver for $\beta_B$, $\bot$ is placed at SN 2. When a batch is sb-delivered successfully, the use of TOB guarantees that all non-faulty nodes will have the same batch with the same sequence number in their log. When all preceding sequence numbers are filled in the log, a request is SMR-DELIVERED to clients. Clients consider their requests delivered when they obtain a quorum of response messages from nodes.

\subsubsection{Multiple Leader Selection}
\label{sec:iss:mls}
For each epoch, ISS selects multiple leaders using a leader selection (LE) policy. The LE policy can be customized so long as it guarantees liveness: each bucket will be assigned to a segment with a correct leader infinitely many times in an infinite execution. The ISS implementation in the paper uses  BFTMencius~\cite{milosevic2013bounded} to keep sufficient correct nodes in the leader set (i.e. the set of all leader nodes). Operating under BFT, BFTMencius uses multiple parallel leaders with its Abortable Timely Announced Broadcast communication primitive to achieve consensus in a bounded delay after GST.

To detect leader failure, each SB instance uses a failure detector that identifies quiet nodes (which may be crashed nodes, or non-crash faults that are identical to crash faults), as introduced by Malkhi and Reiter~\cite{malkhi1997unreliable}. The failure detector guarantees that all correct nodes will eventually suspect a quiet node and that a correct node will not be suspected by other correct nodes after a predetermined amount of time. When a faulty node is detected, ISS removes the node from the leader set and assigns a new leader to the affected SB instance using the leader selection policy.

\subsubsection{Pros and cons}

ISS's greatest contribution is its ability to run multiple consensus instances in parallel, which significantly increases throughput compared to single-leader consensus protocols. ISS also provides faster client responses by allowing client requests to be executed once they are sent, independent of the current epoch being complete. Further, ISS ensures that resources are not wasted processing duplicate requests by sorting each request into only one bucket.

While ISS achieves higher throughput, it introduces additional overhead in consensus latency. For example, client requests must be sent to and hashed at all nodes to ensure they are placed in the appropriate bucket queue. Another drawback is that the multi-leader approach can result in a higher probability that an SB instance will fail when a node becomes faulty. For any leader failure, a new leader must be selected and re-propose uncommitted requests. This can negatively impact overall throughput if the distribution of requests to leader nodes is unbalanced, such that certain node failures cause higher setbacks for the system.
\subsection{DAG-Rider: All You Need is DAG}
\label{sec:dagrider}
\textit{Directed acyclic graph (DAG).}
DAG-based BFT algorithms introduce a novel approach by employing parallel transaction pipelining and constructing directed acyclic graphs on each node. These graphs encapsulate the complete history of message propagation, encompassing both the traversal path and the causal relationships between messages.

One significant advantage of DAG-based BFT systems is their ability to operate without relying on designated leader servers for the transaction distribution process. In traditional leader-based algorithms (e.g., SBFT~\cite{gueta2019sbft} and HotStuff~\cite{yin2019hotstuff}), designated leader servers shoulder the responsibility of transaction dissemination and achieving consensus on those transactions. However, this approach can exert an immense workload on the leader servers, leading to transaction backlogs and system bottlenecks. In contrast, transactions in DAG-based BFT algorithms are pipelined in parallel using the DAG data structure; this is possible as DAG-based algorithms can facilitate a clear separation between the transaction distribution and consensus phase. This results in a substantial increase in throughput, making them a more scalable and efficient BFT consensus solution.

In addition, DAG-based algorithms present a unique fusion of leader-based and leaderless approaches. In the message dissemination phase, no leader is involved, whereas in the consensus phase, a leader is needed for each consensus instance. Instead of queuing messages and being blocked by previous consensus results, nodes can now disseminate and pipeline messages in a non-blocking manner, independent of the ongoing consensus process. The actual consensus for committing the pipelined messages operates as a separate process.

As a result, DAG-based BFT algorithms often demonstrate a clear throughput advantage compared to their counterparts, such as PBFT~\cite{castro1999practical} and HotStuff~\cite{yin2019hotstuff}. Although DAG BFTs suffer from high latency as the final consensus result must wait for the transition pipelining, their throughput advantage makes them a promising choice for scalable and throughput-critical blockchain solutions.

\subsubsection{System model and service properties}
\label{dagrider:sec:systemmodel}
DAG-Rider~\cite{keidar2021all} is an asynchronous Byzantine Atomic Broadcast (BAB) protocol that achieves post-quantum safety with optimal time and message complexity. DAG-rider is comprised of two layers: a round-based structured DAG for reliable message dissemination, and a zero-overhead consensus protocol that allows nodes to independently determine a total order of messages to commit without extra communication overhead.

DAG-Rider uses the Bracha broadcast (see Section~\ref{sec:leaderlesspreliminary}) and tolerates $f$ faulty nodes with at least $3f+1$ total nodes. It ensures liveness by assuming for a \textit{global perfect coin} that allows calling nodes to independently choose a "leader" node (vertex) at an instance of waves (a similar concept to consensus instances) such that all correct nodes choose the same leader. 

\subsubsection{The DAG Pipelining}
Each node independently maintains its own DAG, where the DAG represents the node's interpretation of the message dissemination topology among all nodes. The progression of the DAG occurs in rounds, with each round $r$ building upon the previous one, starting with the formation of the DAG in round $r$ and then adding nodes in round $r+1$. Each node manages an array denoted by $[]$, which stores the DAGs' pertinent information, including the number of vertices ($v$) and the associated round number ($r$). Each vertex within the DAG contains crucial details, such as the message's originator, identified as the "source", the round number, and a block of transactions.

\begin{figure*}[t]
        \centering
        \includegraphics[width=.7\linewidth]{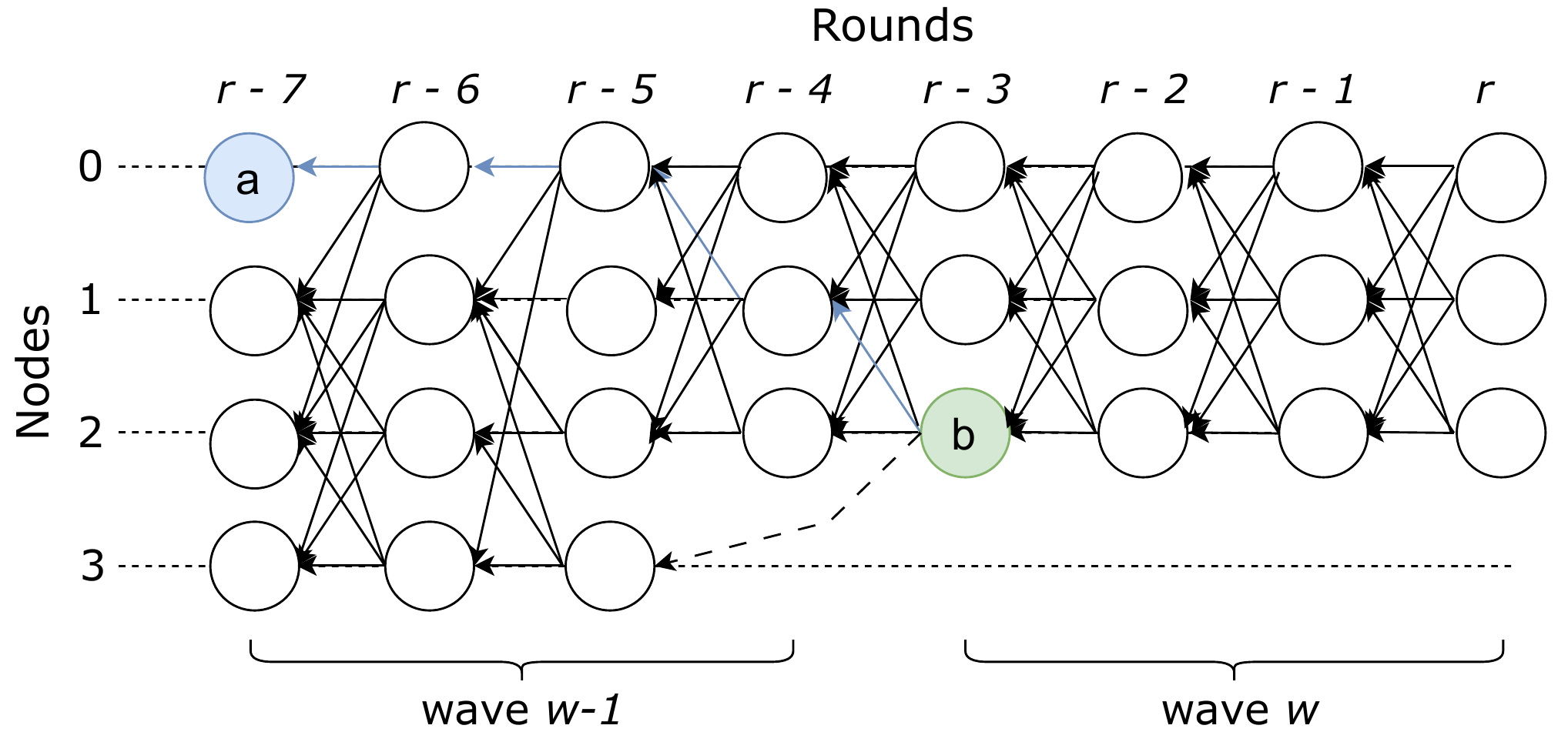}
        \caption{An example of a DAG that contains eight rounds and two waves, where vertices $a$ and $b$ are the leader vertices of waves $w-1$ and $w$, respectively.
        In this DAG, Vertex $b$ has a weak edge connecting it to Node 3, and other links are strong edges that directly connect vertices in adjacent rounds. Since Vertex $a$ does not have sufficient strong edges connecting to it, it is not committed in Wave $w-1$. In contrast, since Vertex $b$ has sufficient strong edges, the transactions stored in both $a$ and $b$ will be committed in Wave $w$, as $a$ has a path linking to $b$.}
        \label{fig:dag1}
\end{figure*}

Figure~\ref{fig:dag1} provides a visual representation of an eight-round DAG on a node. In each round, a vertex is connected to those that disseminate the same message. For instance, in Round $r-6$, the vertex on Node 0 is linked to all nodes in Round $r-7$ because they have all disseminated and stored value $a$.

In addition, DAG-Ridger defines two sets of edges: \textit{strong} and \textit{weak} edges. In a round $r$, strong edges reference the vertices that are directly linked in the previous round (i.e., Round $r-1$), and weak edges reference vertices that are not directly linked in the previous round. In Figure~\ref{fig:dag1}, all the solid arrows are strong edges as they reference connected adjacent vertices. However, the dashed arrow extending from Node 2 in Round $r-3$ (storing Value $b$) forms a weak edge, linking it to Node 3 in Round $r-5$, reflecting the absence of a direct connection as the DAG progresses through rounds.

When clients propose transactions, nodes actively receive vertices, store them in a buffer, and consistently monitor the status of these vertices. The nodes perform periodic checks to ascertain whether the vertices within the buffer have successfully received all their required predecessor vertices (either strong or weak edges). Once the necessary predecessors are included, nodes proceed to integrate the vertices into the DAG array, organizing them based on their respective round numbers. Each attribute of the array can be considered a distinct set that encompasses vertices originating from the corresponding round index.

Upon successfully inserting $2f+1$ vertices from the current round (Round $r$), a node progresses to the subsequent round (Round $r+1$). In this new round, the node generates a new vertex denoted as $v$, which encompasses the essential vertex information: the updated round number, strong edges that connect it to vertices in Round $r$, weak edges extending toward vertices that do not possess any path connecting them to $v$, and the accompanying block of transactions. This vertex is then broadcast to all other replicas within the network.

In the process of constructing DAGs, nodes may, at times, perceive different DAGs due to the protocol's operation within partially synchronous networks. Nevertheless, as each node adheres to the BAB protocol (as detailed in \S~\ref{dagrider:sec:systemmodel}) for vertex dissemination, the DAGs will eventually converge and become identical among non-faulty nodes. DAG-Rider~\cite{keidar2021all} provides rigorous proofs that every vertex broadcast by a correct process is eventually included in the DAG of all other correct processes.

\subsubsection{Consensus in DAGs}
DAG-Rider leverages the previously mentioned perfect coin mechanism, enabling nodes to independently inspect the DAG and deduce which blocks to deliver, as well as their specific order, without necessitating additional coordination among replicas. Following the delivery of vertices in the DAGs across nodes, the system initiates consensus operations aimed at deterministically finalizing the commit of the disseminated transactions.

In this process, DAGs are divided into \textit{waves}, with each wave spanning four consecutive rounds. During each wave, nodes utilize the perfect coin to select a random \textit{leader} vertex in the first round. 
The next step is to deliver all blocks of vertices that either have paths from the leader vertex or are part of the \textit{causal history} of the leader, adhering to a pre-defined deterministic order. To be chosen as a leader for a wave, a vertex must have a crucial property: it must have $2f+1$ strong edges leading to it in the next round. This requirement guarantees that all nodes in the network commit to the same leader every wave. 

Figure~\ref{fig:dag1} is divided into two distinct waves. In the first wave $w-1$, Vertex $a$ serves as the leader vertex. However, in this scenario, the blocks contained in Vertex $a$ cannot be finalized for commitment due to the insufficient presence of strong edges (i.e., fewer than $2f+1$ strong edges). In contrast, in the subsequent wave $w$, Vertex $b$ is elected as the leader vertex. Since Vertex $b$ has sufficient strong edges among the nodes, its blocks will be committed by the end of this wave. Notably, since Vertex $a$ is a part of the causal history of Vertex $b$ (i.e., there exists a path connecting Vertex $a$ to Vertex $b$), Vertex $a$ is also committed alongside Vertex $b$.

\subsubsection{Pros and cons}
The DAG structure employed by DAG-Rider serves a dual purpose of disseminating messages and facilitating voting for agreement. Using the underlying reliable broadcast protocol prevents guarantees that all correct processes eventually have a consistent view of the DAG. The round-based DAG allows for a more stable commit as the expected value of commit is every one and a half waves. 

Compared to leader-based BFT algorithms, DAG-based approaches enable transactions pipelining in parallel. While they obtain high throughput, the latency experiences a significant surge because of the separation between the transaction distribution and consensus. DAGs have a message complexity of $O(n^3\log(n)+nM)$ including the reliable broadcast and a minimum time complexity of $O(5)$ for a strong path. 

\subsection{Narwhal and Tusk: A DAG-based Mempool and Efficient BFT Consensus}
\label{sec:narwhal}
Narwhal~\cite{danezis2022narwhal} uses a DAG-based Mempool to separate transaction dissemination from transaction ordering to achieve high-performance consensus. It offloads reliable transaction dissemination to the Mempool protocol and only relies on consensus to sequence a small amount of metadata, increasing performance significantly. It also introduces Tusk, a BFT consensus protocol that works with Narwhal to achieve high performance even in the presence of faults or asynchrony. 

\subsubsection{System model and service properties}

Narwhal's failure assumption is the same as DAG-Rider (see \S\ref{sec:dagrider}), which tolerates up to $f$ failures with $n=3f+1$ nodes in total. The Mempool works as a key-value store where the value is a block $b$ of transactions and the key is its digest $d$. It provides the unique following service properties.
\begin{itemize}
    \item \textit{Integrity.} It ensures that reading a $d$ also returns the same $b$ on honest parties.

    \item \textit{Block-Availability.} It ensures that reading a $d$ after a successful write of KV-pair $(d, b)$ on a correct party must eventually return $b$.

    \item \textit{Containment.} It ensures that the causal history of a later block always contains that of the earlier block.

    \item \textit{2/3-Causality.} It ensures that the causal history of a block contains at least $2/3$ of the previous blocks.

    \item \textit{1/2-Chain Quality.} It ensures that at least $1/2$ of a causal history is written by honest parties.
\end{itemize}

\subsubsection{The DAG replication protocol}
Narwhal uses a similar method as DAG-Rider for block propagation and creating DAGs; however, it presents a novel implementation of reliable broadcast as opposed to using the expensive Bracha's broadcast protocol. It leverages the inherent DAG structure and introduces \textit{certificate of availability}. The DAG in Narwhal is composed of \textit{blocks}, each comprising a hash signature, a list of transactions, and a collection of \textit{certificates} of availability for the previous round's block. The certificate is a data structure containing the hash of its corresponding block, along with $2f + 1$ signatures from other validators and the round number $r$, acknowledging the successful delivery of the block.

The protocol roughly works as follows:
\begin{enumerate}
\item Validators continually accept client transactions and store them in a transaction list. Additionally, they receive certificates and store them in a separate certificate list.

\item At round $r-1$, if a validator receives $2f+1$ certificates from other validators for the current round, it advances to a new round $r$. Subsequently, it creates a new block containing the current transactions from the transaction list and broadcasts this block to other validators.

\item Upon receiving a block, a validator first validates the block's signature, ensuring it contains a $2f+1$ certificate from the previous round $r-1$. Additionally, the validator ensures that it is the only block from the source validator. If these conditions are met, the validator acknowledges the block by signing the hash with its own signature.

\item Once a validator receives $2f+1$ acknowledge signatures for a specific block it had broadcasted, it creates a certificate of availability for the given block and broadcasts this certificate. The validator then stops broadcasting the original block.
\end{enumerate}

After the dissemination of transactions in the DAG Mempool, Tusk takes charge of managing the consensus process. Tusk chooses a random leader block using the same interpretation in DAG-Rider. Nevertheless, unlike DAG-Rider, Narwhal and Tusk reduce the number of waves used for consensus. They only take a wave of $3$ rounds, and the first and third rounds of two consecutive waves can overlap to allow for a total time complexity of the consensus process of $4.5$ rounds.

\subsubsection{Pros and cons}
Narwhal and Tusk provide a practical implementation as a DAG-based BFT protocol. Similar to DAG-Rider, Narwhal shows a throughput advantage in empirical experiments compared to traditional BFT algorithms, such as PBFT~\cite{castro1999practical} and HotStuff~\cite{yin2019hotstuff}. However, since no order is maintained in the Mempool, Narwhal suffers from a much higher time complexity as transactions that have not met the criteria for a delivery wait for more waves. Thus, transaction consensus often has a high latency because of the waiting time for multiple rounds before the system commits a block.

\section{Toward more robust BFT consensus}
\label{sec:robustness}

While efficient BFT algorithms achieve high performance under normal operation, they may have sacrificed robustness and become vulnerable under a diverse attack vector. This section introduces \emph{robust BFT algorithms} that mark fault tolerance as the first priority, fortifying system robustness by defending diverse attacks under Byzantine failures.
The surveyed algorithms are exemplary approaches in defending against performance attacks (e.g., Aardvark~\cite{clement2009making} in~\S\ref{sec:aardvark}), biased client handling attacks (e.g., Pompe~\cite{zhang2020byzantine} in~\S\ref{sec:pompe}), monopoly attacks (e.g., Spin~\cite{spin2009} in~\S\ref{sec:spin}), and repeated misbehavior attacks (e.g., Prosecutor~\cite{zhang2021prosecutor} in~\S\ref{sec:prosecutor}). This section also includes RBFT~\cite{aublin2013rbft} (in~\S\ref{sec:rbft}), which introduces more redundancy than standard BFT approaches to tolerate corner cases.

\subsection{Aardvark: Making Byzantine fault-tolerant systems tolerate Byzantine failures}
\label{sec:aardvark}

\begin{figure}[t]
    \centering
    \includegraphics[width=0.8\linewidth]{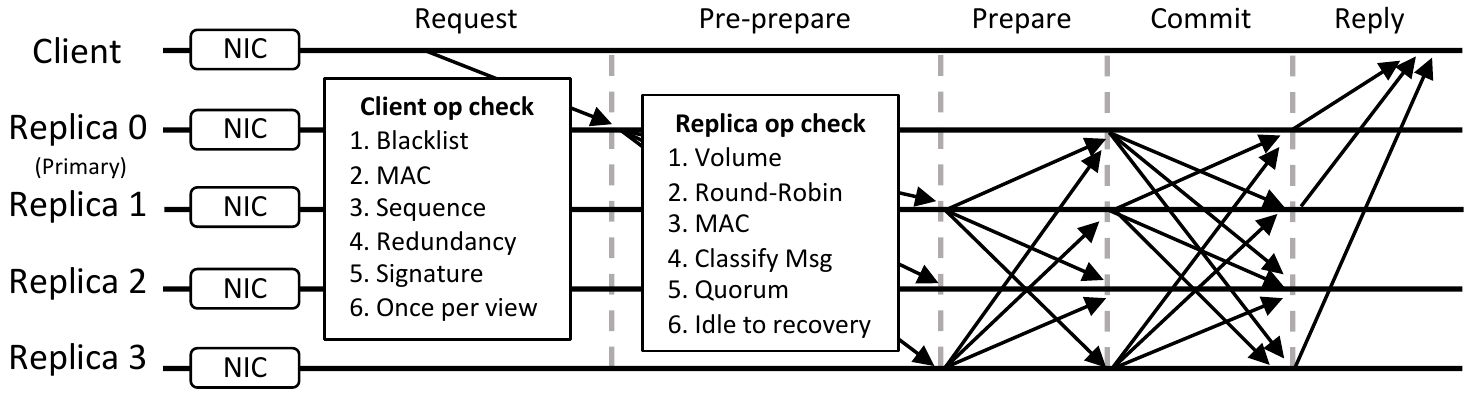}
    \caption{The messaging pattern with robustness improvements under normal operation in Aardvark.}
    \label{fig:aardvark}
\end{figure}

Aardvark~\cite{aadvark2009} aims to improve the robustness of BFT algorithms that achieve high performance in terms of throughput and latency when the system operates in \emph{gracious executions}, such as PBFT~\cite{castro1999practical}, Zyzzyva~\cite{kotla2007zyzzyva}, 700BFT~\cite{guerraoui2010next}, and Scrooge~\cite{serafini2008scrooge}. In gracious executions, the network is synchronous, and all clients and servers are correct. However, the efficient BFT protocols are vulnerable during \emph{uncivil executions}, in which the network latency is bounded and up to $f$ servers may become Byzantine faulty with unlimited Byzantine clients. In this case, a series of meticulously designed attacks can wreck the performance of these BFT algorithms, and the system availability decreases significantly (even down to zero). For example, when clients use an inconsistent authenticator on requests, PBFT~\cite{castro1999practical} fails to make progress and can result in repeated view changes. 

\subsubsection{System model and service properties} 
Aardvark uses the same fault-tolerance and network synchrony assumption as PBFT: it tolerates $f$ fault replicas out of a total of $3f+1$ replicas; it ensures safety in asynchrony but requires bounded message delay to obtain liveness. Aardvark argues that the first principle to design BFT algorithms should be the ability to tolerate Byzantine failures and defend against malicious attacks; during uncivil executions, the performance degradation should remain at a reasonable degree. Aiming to improve BFT robustness, Aardvark's improvements are threefold: stronger message authentication (signed client requests), independent communication (resource isolation), and adaptive leadership rotation (regular view changes).

\subsubsection{Robustness in log replication}
Aardvark requires clients to use digital signatures to authenticate their requests. Unlike the message format of PBFT in Table~\ref{tab:PBFT_messaging_formats}, a client request is specifically authenticated between the requesting client (denoted by $c$) and the receiving replica (denoted by $r$). It has the format of $\langle \langle \textsc{request}, op, timestamp, clientId \rangle_{\sigma_c}, clientId\rangle_{\mu_{c,r}}$, where the operation and timestamp are signed by signature $\sigma_c$; the signed message is then authenticated by a MAC (message authentication code) $\mu_{c,r}$ for recipient $r$. In addition, Aardvark requires communication among replicas to use separate network interface controllers (NICs) and wires (shown in Figure~\ref{fig:aardvark}). This separation enables a more secure communication channel, reducing interference from faulty servers. Replicas also separate message queues for receiving requests from external clients and internal replicas. The isolation of message queues prevents replicas from being overwhelmed by a spike of client requests. 

The workflow of the replication protocol of Aardvark is similar to that of PBFT. However, with signed client requests and resource isolation, Aardvark takes more steps to verify both client and replica messages, aiming to reduce the potential impairment from faulty clients and replicas. To verify a client message, a replica, in the request phase, checks that \textcircled{1} the client is not blacklisted; \textcircled{2} the authentication MAC ($\mu_{c,r}$) is valid; \textcircled{3} the timestamp is incremented by one as the previously received one; and \textcircled{4} the request has not been processed. Then, \textcircled{5} the replica verifies signature $\sigma_{c}$; if $\sigma_{c}$ is invalid, the replica blacklists this client. Finally, if the request is verified in a previous view but not processed, \textcircled{6} the replica continues the replication protocol, which allows the system to commit cross-view requests.

Messages circulated between two replicas also undergo a checklist process. In particular, when replica $i$ receives a message from replica $j$, replica $i$ \textcircled{1} takes a volume check: if replica $j$ sends more messages than expected, replica $i$ blacklists replica $j$. Then, replica $i$ \textcircled{2} picks received messages to process consensus in a round-robin manner and discards incoming messages if buffers are full. Next, replica $i$ \textcircled{3} verifies the message's MAC, \textcircled{4} processes the message according to its type, and \textcircled{5} adds it to the corresponding quorum. After a complete quorum (with a size of $2f+1$) is formed, the consensus process continues. If replica $i$ is not processing any messages for forming consensus (i.e., in the idle mode), replica $i$ \textcircled{6} starts to process catchup messages from other servers in the recovery mode.

\subsubsection{Robustness in view changes}
In PBFT, a primary is considered correct if consensus can be achieved in time; i.e., backups commit a proposed request before their timers expire. This primary may stay in this position forever and dominate the replication phase. This scenario may result in low performance if the primary deliberately slows the process of leading the consensus or colludes with faulty clients. 
Although Aardvark utilizes the same view-change protocol as PBFT~\cite{castro1999practical} (shown in Figure~\ref{fig:pbft-view-change}), to prevent any replica from dictating the replication process, Aardvark imposes two expectations on a correct primary: \textcircled{1} sufficiently and timely issuing \textsc{Pre-prepare} messages and \textcircled{2} maintaining sustained throughput.
If the primary fails either of the two expectations, Aardvark imposes view changes regardless of the primary's correctness.

The former expectation requires the primary to timely and fairly process client requests. Aardvark manages \textsc{Pre-prepare} messages as heartbeats between the primary and backups. A backup starts a timer in a view, say view $v$, and resets it upon receiving a valid \textsc{Pre-prepare} message from the primary; otherwise, the timer keeps counting down. Thus, if the primary cannot issue \textsc{Pre-prepare} messages in time, the backup's timer expires; the backup considers the primary to be faulty and invokes a view change to enter view $v+1$. In addition, backups require fairness of client request ordering. A backup monitors a client request if the request has not been pre-prepared and relays it to the primary. If the primary cannot issue \textsc{Pre-prepare} for this request after another two requests are pre-prepared, the backup considers the primary to be faulty as the primary's behavior is unfair, with a probability of trying to block requests from specific clients. Then, the backup initiates a view change to enter the next view. 

The latter expectation is enforced by monitoring the primary's throughput over $n$ previous views. Aardvark uses periodic checkpoint messages to calculate a primary's averaged throughput. The primary of view $v$ is expected to maintain at least $90\%$ of the throughput achieved in its $n$ prior views. If the primary fails to accomplish this goal, backups consider the primary to be faulty and initiate a new view change to enter view $v+1$.

\subsubsection{Pros and cons}
Aardvark addresses the robustness problem of PBFT-like consensus algorithms, especially focusing on tolerating faulty clients. 
The improvements made by Aardvark are generally focused on requiring signed client messages, isolating resources, and performing regular view changes; thus, compared with its baseline protocols, Aardvark obtains a lower throughput deduction during uncivil executions than during gracious executions.

On the other hand, the strict expectations on a correct primary to maintain at least $90\%$ of the throughput achieved in the last $n$ views may bring unnecessary burden on correct replicas and benefit a faulty primary~\cite{aublin2013rbft}. When the workload increases, a faulty primary can block incoming client requests and select only $90\%$ of them to process. 
This behavior may result in more severe performance degradation when the workload is dynamic. Additionally, since correct primaries can be mistakenly replaced with the system making no progress, Aardvark may impose unnecessary view changes, thereby impeding higher throughput performance. 
\subsection{Pompe: Byzantine ordered consensus without Byzantine oligarchy}
\label{sec:pompe}
\begin{figure}
    \centering
    \includegraphics[width=0.8\linewidth]{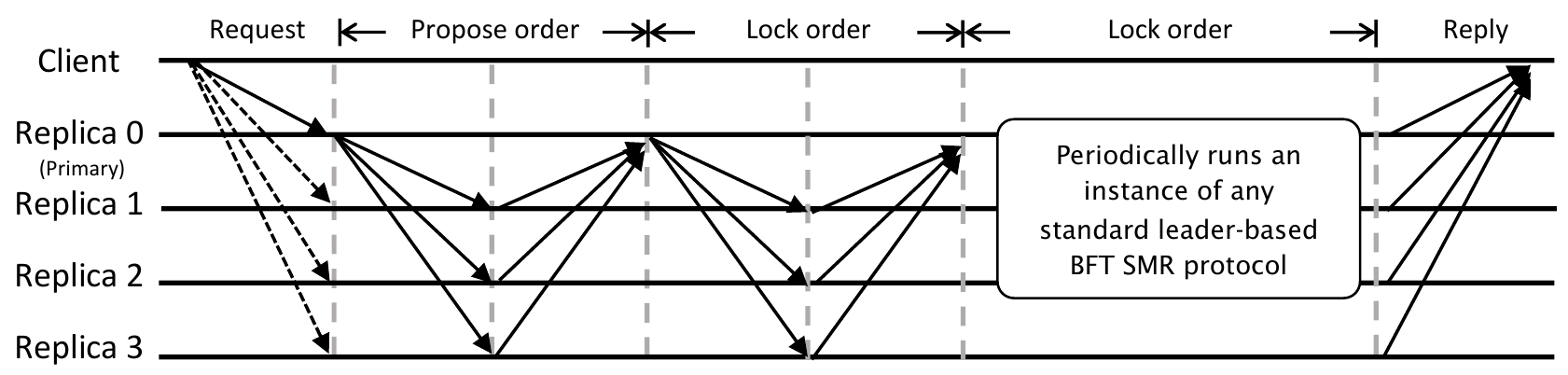}
    \caption{The message-passing workflow of determining an order that avoids primary manipulation in Pompe.}
    \label{fig:pompe-normal-op}
\end{figure}

Pompē~\cite{zhang2020byzantine} is an ordering protocol built on top of standard consensus algorithms (e.g., SBFT~\cite{gueta2019sbft} and HotStuff~\cite{yin2019hotstuff}). It prevents Byzantine replicas from dictating the order of commands by democratically collecting preferred orders from a quorum of replicas. After an order of a command is decided, Pompē relies on an applied consensus algorithm to perform consensus for a batch of ordered commands in a periodic manner.

\subsubsection{System model and service properties}

Pompē tolerates $f$ faulty replicas out of a total of $3f+1$ replicas. It has a unique service property that can address the ordering of operations/commands since the ordering of operations can have significant financial implications~\cite{daian2020flash}. In addition to the traditional specification of BFT SMR correctness (i.e., safety and liveness), Pompē introduces a new property: \emph{Byzantine ordered consensus}. This property assigns each operation with an ordering indicator, which shows a replica's preference to order the operation. By coordinating indicators, Pompē avoids the ordering manipulation of operations against Byzantine replicas (i.e., Byzantine oligarchy) that disrespect the ordering indicators from correct replicas. For example, although it can ensure safety and liveness, a Byzantine primary can always deliberately put requests from client $c_i$ before requests from client $c_j$ when their requests invoke the same operation.

\subsubsection{The replication protocol}
Pompē has two phases in the replication protocol: \textcircled{1} the \emph{ordering phase} and \textcircled{2} \emph{consensus phase}. The ordering phase secures a democratic order of a proposed request. Unlike traditional BFT algorithms, in which the order is single-handedly decided by the primary, in Pompē, the order is a selection among $2f+1$ preferred orders from different replicas. Specifically, as shown in Figure~\ref{fig:pompe-normal-op}, the ordering phase has two steps: the \emph{propose order} and \emph{lock order} steps. In the first step, a replica (say replica $i$), acting as a proposer, broadcasts a \textsc{RequestTS} message in the format of $\langle \textsc{RequestTS}, c \rangle_{\sigma_i}$ to solicit preferred timestamps, which are used as the ordering indicator, for command $c$ from all other replicas. Then, a replica (say replica $j$) replies to the proposer with $\langle \textsc{ResponseTS}, c, ts \rangle_{\sigma_j}$, where $ts$ is its preferred timestamp. After the first step, the proposer collects \textsc{ResponseTS} messages to a set $\mathcal{T}$ until a quorum is formed. 
Then, the second step starts; the proposer picks the median timestamp, denoted by $ts_m$, from $\mathcal{T}$, and assigns $ts_m$ as the order of the command. Next, the proposer broadcasts the order in the format of $\langle \textsc{Sequence}, ts_m, c, T \rangle_{\sigma_i}$.
Finally, replica $j$ verifies and accepts the received order if the received order should be higher than any previously accepted orders. If the received order is accepted, replica $j$ replies to the proposer with an acknowledgment in the format of $\langle \textsc{SequenceResponce}, ack, h \rangle_{\sigma_j}$; otherwise, it replies $\langle \textsc{SequenceResponce}, nack, h \rangle_{\sigma_j}$, where $h$ is the cryptographic hash of the received \textsc{Sequence} message from the proposer. At the end of the ordering phase, Pompē obtains an order for a specific command such that the order is not manipulated by Byzantine oligarchy. Since there are at most $f$ Byzantine replicas, in a quorum with a size of $2f+1$, the median is the both upper- and lower-bounded timestamps from correct replicas.

Pompē relies on an applied consensus algorithm to obtain consensus for ordered commands. For example, if Pompē uses HotStuff~\cite{yin2019hotstuff} as its consensus algorithm, it periodically executes consensus in a predefined interval. During an interval, Pompē packs ordered commands into a batch, and the consensus phase is used to periodically commit a batch of commands. With the ordering and consensus phases, Pompē achieves Byzantine ordered consensus, which adds ordering fairness to linearizability.

\subsubsection{The view-change protocol}
Pompē does not have a new design for view changes. When a primary fails, both the ordering and consensus phases cannot proceed.
Pompē relies on its applied consensus algorithm to detect primary failures and then initiate a view change.

\subsubsection{Pros and cons}
Pompē extends the traditional correctness discussion of BFT SMR from focusing only on safety and liveness to the ordering of operations. This new property may have a significant impact when the BFT SMR service is used in financial applications~\cite{daian2020flash}. In Pompē, the ordering of operations is democratically decided by a quorum of replicas instead of being determined by a single primary. Pompē separates consensus into two major phases as ordering and consensus and majorly discusses the ordering phase. 

Yet, Pompē does not design a specific consensus algorithm but uses any standard consensus algorithm to perform the second phase. However, since many consensus algorithms have meticulously designed view-change protocols, it is unclear how safety and/or liveness can be ensured when these consensus algorithms are applied to perform the ordering service. For example, PBFT's design allows valid cross-view requests to be committed; if PBFT applies to Pompē as a consensus algorithm, it may disable this feature because of potential safety violations.

\subsection{Spin: Byzantine fault tolerance with a spinning primary}
\label{sec:spin}

Spinning~\cite{spin2009} addresses the problem of performance degradation in leader-based BFT algorithms under faulty leaders. PBFT~\cite{castro1999practical} and its variants~\cite{kotla2007zyzzyva, bessani2014state, aadvark2009} use a stable primary to achieve consensus, and backups detect primary failures using timeouts. The primary remains unchanged as long as no failures are reported by backups. The timeout value is often set much larger than the expected completion time of a consensus instance (e.g., a timeout is set to $500$~ms when consensus can be achieved within $50$~ms).
However, this stable leadership design is vulnerable to \emph{performance attacks} where a faulty primary deliberately slows down transaction processing in a delay that does not trigger a backup timeout. To mitigate this problem, Spinning rotates primaries for each client request, which avoids leadership monopoly from specific replicas. Since faulty primaries can cause sustained performance degradation, it also uses a blacklist to prevent $f$ suspected faulty replicas from becoming a primary.

\begin{figure*}[t]
    \begin{subfigure}{0.74\textwidth}
        \centering
        \includegraphics[width=\linewidth]{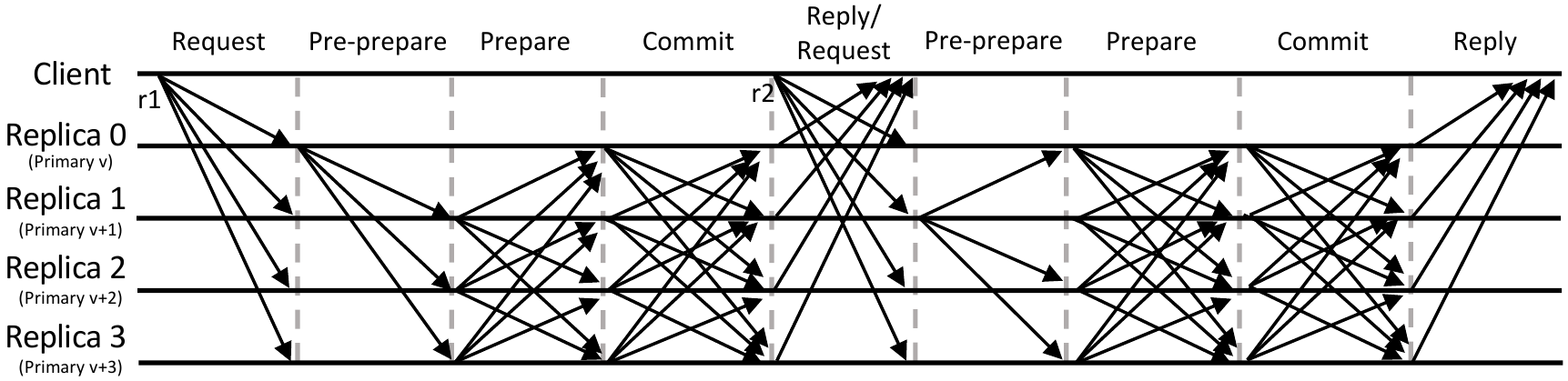}
        \caption{Normal operation of consensus instances for committing two requests.}
        \label{fig:spin-normal-operation}
    \end{subfigure}
    \hfill
    \begin{subfigure}{0.25\textwidth}
        \centering
        \includegraphics[width=0.85\linewidth]{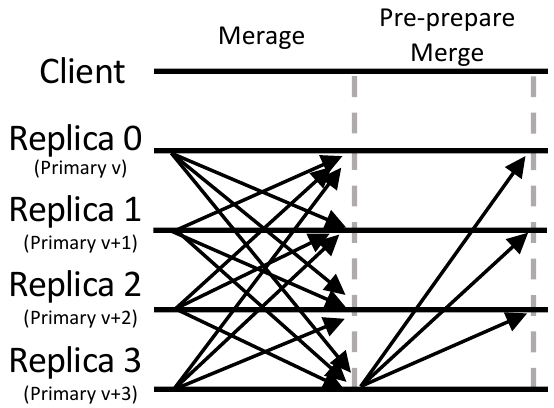}
        \caption{Merge operation upon primary failures.}
        \label{fig:spin-merge}
    \end{subfigure}
    \caption{The message-passing workflow of normal operation and merge operation in Spin.}
\end{figure*}

\subsubsection{System model and service properties}
Spinning assumes a partially synchronous network for maintaining liveness while safety does not require network synchrony. Similar to the fault-tolerant model in PBFT~\cite{castro1999practical}, Spinning tolerates $f$ failures in a total of $3f+1$ replicas. In addition, standard public/private key encryption is used on messages. 
 
\subsubsection{Robustness in log replication}
Spinning's normal operation for a consensus instance follows PBFT's communication pattern that includes \textsc{pre-prepare}, \textsc{prepare}, and \textsc{commit} phases, with $2f+1$ \textsc{commit} messages needed for operation execution (shown in Figure~\ref{fig:spin-normal-operation}).
Compared with PBFT, which changes a primary when it is suspected of being faulty, Spinning changes a primary whenever it defines the order of a batch of requests. For example, in Figure~\ref{fig:spin-normal-operation}, when a client broadcasts a request ($r1$), replica $R_0$, the primary of view $v$, starts a consensus instance for committing $r1$; when the client proposes another request ($r2$), the consensus is conducted by the primary ($R_1$) of the next view, $v+1$. Since requests are handled in different views, Spinning does not need to assign sequence numbers to requests but uses the view number instead.
After a client broadcasts a request, if a replica receives a \textsc{pre-prepare} message from the primary in view $v' > v$ for current view $v$, it buffers the message until view $v'$ becomes valid; $v$ is set when the replica receives at least $f+1$ \textsc{prepare} messages from distinct replicas with $v$ as the current view number. If a replica cannot receive enough \textsc{prepare} messages, it uses the last accepted view number $v_{last}$. When the client receives $f+1$ replies from distinct replicas, it accepts the result in the reply. 

Replicas use timeouts to limit the completion time of a consensus instance. If a replica cannot collect enough \textsc{commit} messages in time, it triggers a timeout and sends a \textsc{merge} message to all replicas. Unlike PBFT, the merge operation is not designed to change views; it aims to determine which requests from the previous view can be accepted and executed by merging information from all replicas and proceeding to a new view (shown in Figure~\ref{fig:spin-merge}). The merge operation first begins when replicas trigger a timeout or have received $f+1$ \textsc{merge} messages from distinct replicas.
A merge message contains a set, $P$, that contains requests that a replica has received a \textsc{pre-prepare} message paired with $2f+1$ \textsc{prepare} messages. That is, set $P$ shows requests that have been prepared and can proceed with their commit operation in the next view. Next, after receiving $2f+1$ merge messages, the replica of the new view enters the \emph{pre-prepare merge} phase ($R_3$ in Figure~\ref{fig:spin-merge}). It broadcasts a \textsc{pre-prepare-merge} message that contains a vector of digests of requests received from valid $P$ sets. Finally, replicas change their state back to \emph{normal} after a received \textsc{pre-prepare-merge} message is verified, and normal operation resumes.

\subsubsection{Robustness in view changes}

Since leadership is rotated for each request, Spinning does not require an explicit view change protocol when a primary failure is detected. However, under $f$ faulty replicas in a total of $n=3f+1$ replicas, frequent leadership rotation can significantly decrease throughput, as no request can be committed under a faulty primary, with a drop in throughput of $33\%$ in theory ($\frac{f}{3f+1}$). To mitigate this effect, Spinning uses a blacklist to exclude misbehaved replicas when assigning the primary role. The list has a maximum capacity of $f$, where the oldest are removed if full, and is maintained by all correct replicas using \textsc{pre-prepare-merge} messages. 
When a faulty primary triggers a merge operation, the failure is confirmed when a \textsc{pre-prepare-merge} message is sent in a view; replicas then put this faulty primary into the blacklist. 
Since a replica can be wrongly judged faulty because of network conditions, when the blacklist size exceeds $f$, the oldest replica is removed in a FIFO manner. Replicas on the blacklist  can still participate in consensus processes; they are excluded from the primary role only when leadership is rotated for handling new incoming client requests. 

\subsubsection{Pros and cons}
Equipped with the leadership-rotation-per-request mechanism, compared with efficient BFT algorithms (in~\S~\ref{sec:efficiency}), Spinning becomes more robust under malicious attacks targeting the primary replicas. Since it amortizes the coordination workload among all replicas, it also mitigates the performance bottleneck problem in algorithms that use only a single primary.

Nevertheless, leadership rotation incurs extra message-passing when failures do not occur on a primary. The merge operation has a message complexity of $O(n^2)$ and is triggered on every failed replica. This extra messaging cost has a negative impact on system performance whenever a replica fails in the system.

\begin{figure*}[t]
    \begin{subfigure}{0.49\textwidth}
        \centering
        \includegraphics[width=0.9\linewidth]{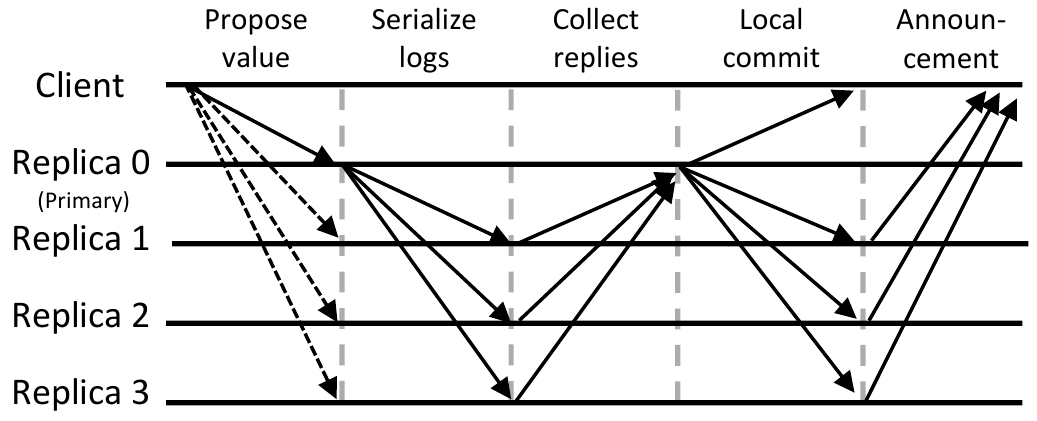}
        \caption{Replication under normal operation.}
        \label{fig:prosecutor-normal-operation}
    \end{subfigure}
    \hfill
    \begin{subfigure}{0.49\textwidth}
        \centering
        \includegraphics[width=0.9\linewidth]{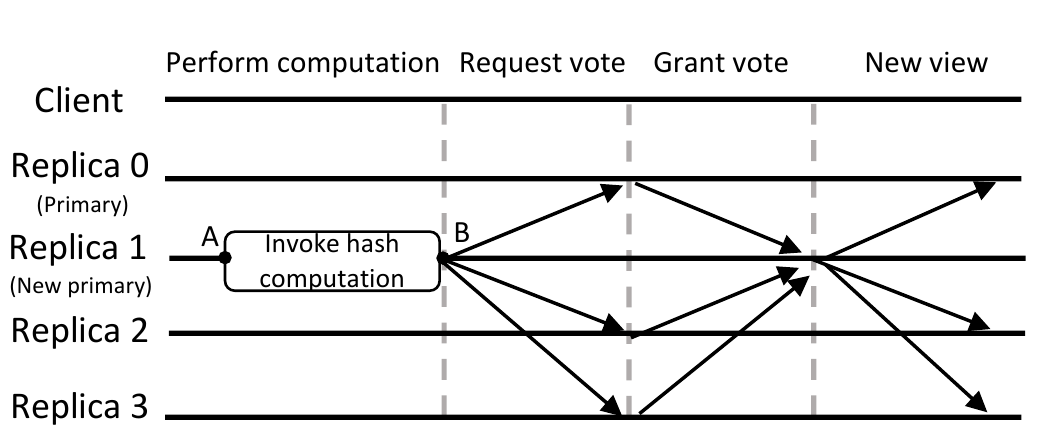}
        \caption{View changes with punitive computation.}
        \label{fig:prosecutor-view-change}
    \end{subfigure}
    \caption{The replication and view-change protocols in Prosecutor.}
\end{figure*}

\subsection{Prosecutor: Behavior-aware penalization against Byzantine attacks}
\label{sec:prosecutor}

\subsubsection{System model and service properties}
Prosecutor~\cite{zhang2021prosecutor} assumes a partially synchronous network and tolerates $f$ Byzantine failures out of a total of $3f+1$ replicas. Compared with PBFT-like algorithms, Prosecutor has a different perspective on view changes: it allows server replicas to actively claim to be the new primary and transition to an intermediate state called \emph{candidate}, as opposed to rotating leadership on a predefined schedule. To suppress Byzantine servers from being the primary, Prosecutor imposes computation penalties on candidates; the computation difficulty changes based on the candidate's past behavior. The more failures a candidate has exhibited, the greater computation penalty the candidate must incur for becoming the new primary.

\subsubsection{The replication protocol}
Prosecutor efficiently replicates client requests, with a message complexity of $O(n)$. Similar to HotStuff's replication protocol, Prosecutor utilizes threshold signatures to reduce the package size in primary-backup communication. Specifically, the replication process has five phases. First, the client broadcasts requests to all replicas. Then, the primary assigns an order to the request and sends it to all other replicas. Next, replicas verify the primary's message to ensure that the order has not been used for other requests. Then, replicas partially sign an endorsement and reply to the leader, indicating that they agree to commit the request in the assigned order. After collecting $2f+1$ replies from different replicas, the leader considers the request to be committed and issues a commit instruction signed by its threshold signatures. Finally, replicas examine the commit instruction and announce to the client that the request has been locally committed. 

\subsubsection{The view-change protocol}
Prosecutor's view-change protocol, associated with computation penalties, is the most unique feature compared with those of other state-of-the-art BFT algorithms. Generally, a view-change process has two major steps: performing computations and voting (shown in Figure~\ref{fig:prosecutor-view-change}). Specifically, if a backup's timer expires, the backup invokes an algorithm, similar to Proof-of-Work~\cite{nakamoto2008bitcoin}, to perform hash computations. 
\textcircled{1} The backup prepares a proof window that contains an array of values starting at the first uncommitted value and ending at the last committed value. Then, the backup combines its proof window with a randomly generated string called \emph{nonce}. 
\textcircled{2} The backup computes the hash of the combination to obtain a hash result. Prosecutor imposes a \emph{threshold} requirement for the result. The threshold is an integer that indicates the number of identical and consecutive bytes a hash result should prefix. \textcircled{3} If the hash result satisfies the threshold requirement, the algorithm terminates; otherwise, the backup repeatedly changes the nonce until obtaining a qualified result. After a qualified result is acquired, \textcircled{4} the backup becomes a candidate, broadcasts a \textsc{VoteMe} message including the new view number and hash result, starts a new timer, votes for itself, and waits for votes from the other replicas. If the candidate collects $2f$ votes before its timer expires, \textcircled{5} it becomes the primary in the new view and broadcasts a \textsc{New-View} message including the collected votes, showing the legitimacy of leadership.

Replicas increment the threshold value of a requesting candidate after receiving its \textsc{VoteMe} message. Thus, the more requests a candidate has initiated, the more computation the candidate must perform for the next leadership competition. Moreover, replicas decrement a candidate's threshold if the candidate becomes the primary and successfully conducts $k$ consensus in a new view, where $k$ is a predefined parameter of the minimum number of expected transactions committed under correct leadership. Therefore, if a replica does not fulfil the primary duty after being elected, its threshold will not be decreased; this penalization regime entices replicas to operate correctly in order to avoid performing computation work.

\subsubsection{Pros and cons}
Prosecutor penalizes suspected faulty replicas while achieving efficient consensus in terms of linear message complexity. If Byzantine servers launch attacks aiming to usurp correct leadership, they will be penalized and forced to perform exponentially growing computational work. Thus, Byzantine servers are gradually suppressed and marginalized from leadership competition. Compared with PBFT-like view change protocols, such as PBFT~\cite{castro1999practical} and HotStuff~\cite{yin2019hotstuff}, Prosecutor can avoid sustained performance degradation when faulty servers repeatedly launch attacks. However, the PoW-like penalization may become less efficient if faulty servers have a strong computation capability; in this case, Prosecutor may suffer from a long period without a correct leader before Byzantine servers exhaust their computation capability.
\subsection{RBFT: Redundant Byzantine fault tolerance}
\label{sec:rbft}
\begin{figure}[t]
    \centering
    \includegraphics[width=0.8\linewidth]{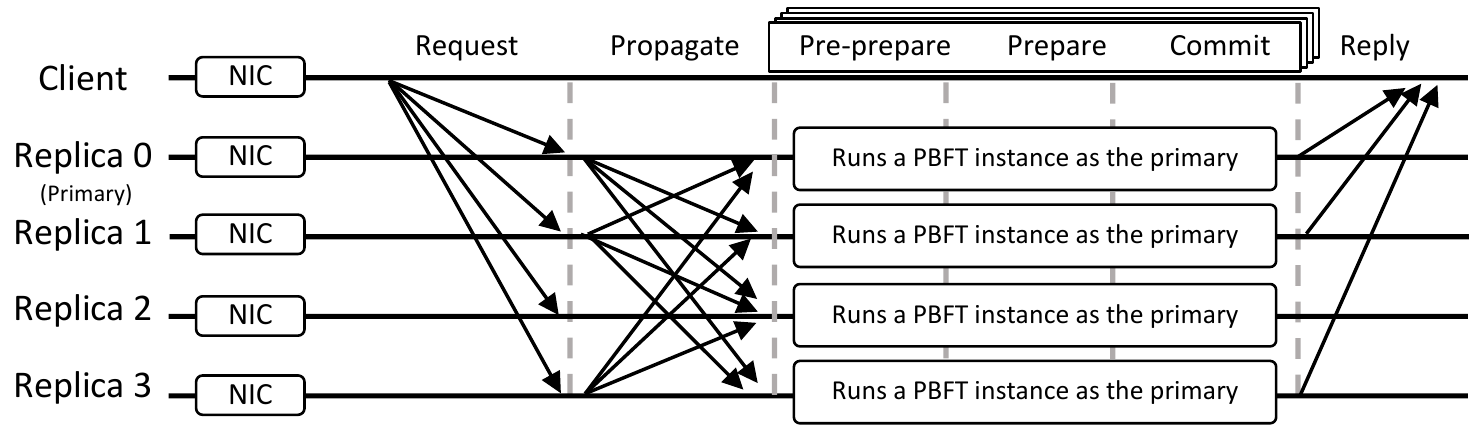}
    \caption{The message-passing workflow of normal operation in RBFT. Replicas run a PBFT consensus instance in parallel after the propagate phase.}
    \label{fig:rbft}
\end{figure}

\subsubsection{System model and service properties}
RBFT~\cite{aublin2013rbft} suggests that some methods aiming to improve the robustness of fault tolerance, such as Prime, Aardvark, and Spin, leave the door open for performance degradation in numerous corner cases. These corner cases often target a primary, which may result in single points of failure, thereby reducing system performance. To tackle this challenge, RBFT introduces more redundancy in the consensus process: RBFT executes multiple instances of the PBFT protocol in parallel, and each replica plays the primary role in the corresponding instance. Among these parallelly executing instances, one instance assumes a master role and takes the primary duty to lead the consensus, whereas other instances assume a backup role to monitor the processing speed of the master, preventing the master from deliberately slowing down the consensus process. Similar to PBFT's failure assumption, RBFT tolerates $f$ faulty replicas out of a total of $3f+1$ replicas.

\subsubsection{Robustness in replication}
The replication process has four major phases, the first of which is the \emph{request} phase. In this phase, a client broadcasts a request to all replicas. In contrast to PBFT, RBFT requires clients to broadcast their requests to all replicas, even if the master replica is correct and responsive. 
After checking the signature of the request, replicas broadcast the request to all other replicas in the \emph{propagate} phase. 
Since only $f$ out of $3f+1$ replicas are faulty, a correct replica is able to eventually receive at least $f+1$ \textsc{propagate} messages. 
Then, the replica starts its local instance of a PBFT consensus, which contains the \emph{pre-prepare}, \emph{prepare}, and \emph{commit} subphases. Similar to the end of PBFT's commit phase, if $2f+1$ \textsc{commit} messages are collected from different replicas in the same instance, the replica in RBFT commits the request. Finally, the replica executes the request and sends a reply  to the requesting client in the reply phase.

The redundancy of multiple concurrently running instances is used to monitor the master instance's processing speed to detect whether the master is faulty. Each instance sends periodic messages to inquire about the number of requests the other servers have processed and then calculates the average processing speed. If the master's processing speed falls below the average by a threshold (say $t_i$), the instance considers the master instance to be faulty and initiates instance changes (similar to view changes). Additionally, each instance tracks the processing time of each request's consensus and calculates the average latency for each client. If the processing time for some clients deviates from the average by a threshold (say $t_c$), the master instance is considered unfair in leading the consensus. Consequently, backup instances initiate instance changes to enter the next view.

\subsubsection{Robustness in instance changes}
 The instance-change protocol operates in a similar way as the view-change protocol in PBFT to replace a faulty primary. Each instance maintains a counter $c$; if the master instance does not satisfy the two thresholds (i.e., $t_i$ and $t_c$), a backup instance increments $c$ and broadcasts an \textsc{instance-change} message in the format of $\langle \textsc{instance-change}, c, i \rangle_{\sigma_i}$, in which $i$ is the instance ID. After receiving this message, other instances check the currently monitored performance of the master instance if the received $c$ is not less than their own counters. If the performance does not satisfy the two threshold criteria, these instances support the instance change and broadcast their \textsc{instance-change} messages. Upon receiving $2f+1$ \textsc{instance-change} messages with the same counter from different instances, a new master instance is selected, and the system enters the next view. 

\subsubsection{Pros and cons}
RBFT observes that some previously published robust BFT algorithms, such as Prime, Aardvark and Spin, have flaws in corner cases. These cases often target the primary and can result in dramatic throughput degradation. RBFT solves this problem by using multiple concurrently running instances to monitor the performance of the primary (master instance). 

Nevertheless, RBFT introduces more operating costs owing to more redundancy; this imposes a massive additional messaging cost on the system compared with PBFT-like algorithms, the message complexity of which (in the order of $O(n^2)$) hinders them from being used in large-scale practical applications. Since each replica runs a PBFT-like instance after the propagate phase, to commit $|M|$ requests, the message complexity of RBFT is $O(|M|(n+n^2+n(n+n^2+n^2)+n)) = O(2|M|(n^3+n^2+n))$. In addition, RBFT may still suffer from unnecessary view changes that mistakenly replace a correct master instance. In practice, especially in geo-distributed applications, the communication latency between clients and different replicas can be significantly different; when the difference causes a correct master to fail to meet the fairness requirement, backups replace the master by initiating unnecessary instance changes.

\section{Toward more available BFT consensus}
\label{sec:availability}

In addition to the efficient and robust BFT algorithms, there has been significant research focusing on addressing the availability problem in BFT consensus. Specifically, researchers have developed asynchronous BFT protocols that are capable of operating in scenarios with varying levels of synchrony. Many leader-based algorithms are susceptible to single points of failure, particularly when targeting the primary server. In such cases, the system must first detect a primary failure and then invoke a view-change protocol to select a new primary. Moreover, primary servers that handle interactions with both clients and other servers often experience heavy workloads, potentially becoming bottlenecks for the entire system.

This section delves into the approaches that offer increased availability, allowing BFT consensus to function with limited synchrony. These more available BFT algorithms predominantly fall into the category of leaderless algorithms. We first introduce the basic components in \S\ref{sec:leaderlesspreliminary} and DBFT~\cite{crain2017dbft} in \S\ref{sec:dbft}, Honeybadger~\cite{miller2016honey} in \S\ref{sec:honeybadger}, and the BEAT family~\cite{duan2018beat} in \S\ref{sec:beats}.

\subsection{Preliminary fundamental models}
\label{sec:leaderlesspreliminary}
Before delving into each algorithm, we introduce two fundamental building blocks that facilitate the design and implementation of leaderless BFT algorithms: the reliable broadcast (i.e., Bracha's RBC), introduced by Bracha et al.~\cite{bracha1987asynchronous}, and the binary Byzantine agreement (i.e., BA), introduced by Moustefaoui et al.~\cite{mostefaoui2014signature}.

\begin{figure*}[t]

    \begin{subfigure}{0.49\textwidth}
        \centering
        \includegraphics[width=\linewidth]{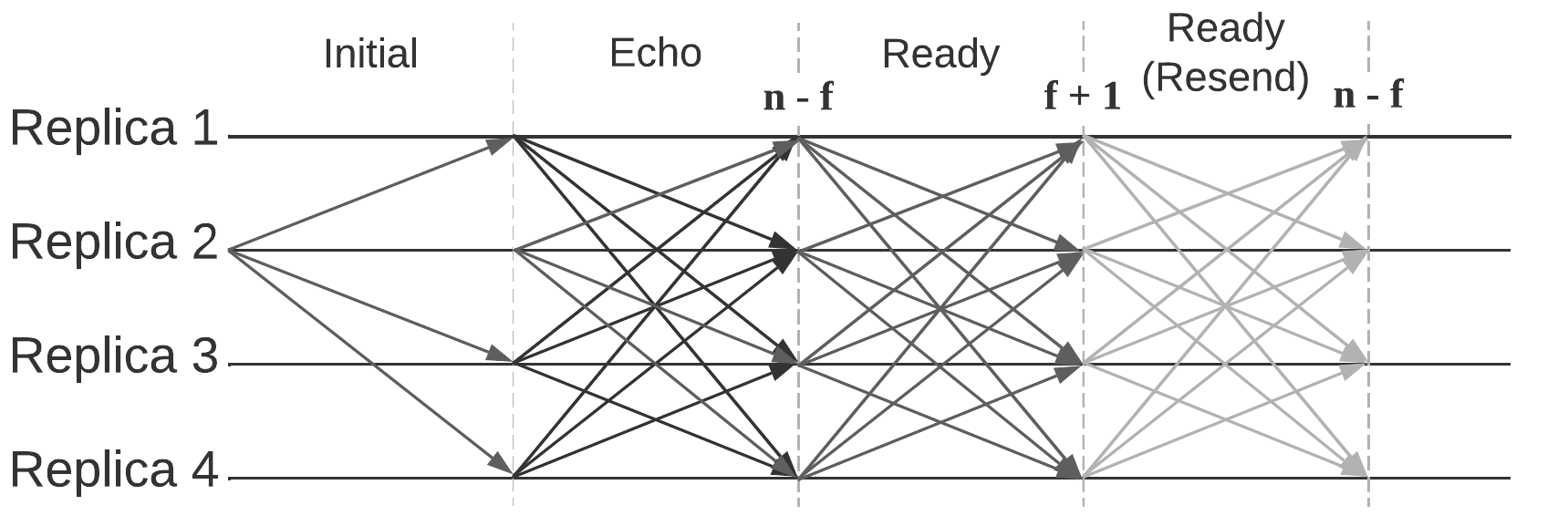}
        \caption{The messaging pattern under normal operation.}
        \label{fig:rbc-regular}
    \end{subfigure}
    \hfill
    \begin{subfigure}{0.49\textwidth}
        \centering
        \includegraphics[width=\linewidth]{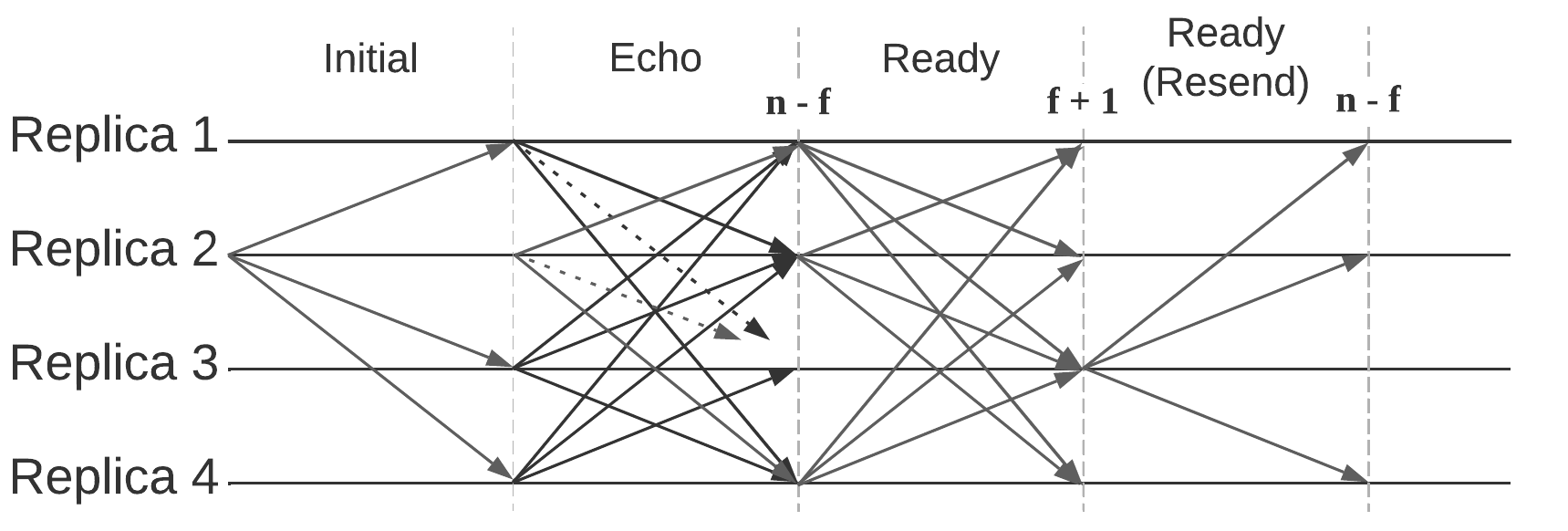}
        \caption{An example of sending the \textsc{ready} message.}
        \label{fig:rbc-example}
    \end{subfigure}
    \caption{The messaging pattern of Bracha's RBC where $n=4$ and $f=1$.}
    \label{fig:rbc}

\end{figure*}

\subsubsection{Reliable broadcast} \label{sec:rbc}
This protocol achieves agreement in asynchronous networks~\cite{bracha1987asynchronous}. It contains four major steps (shown in Figure~\ref{fig:rbc-regular}).

\begin{enumerate}
    \item First, the primary (replica $R_2$) proposes a value $v$ by broadcasting an \msgtype{initial} message.
    
    \item When a replica (including the primary) receives the \msgtype{initial} message, it broadcasts an \msgtype{echo} message with $v$.
    
    \item A replica broadcasts a \msgtype{ready} message with $v$ if it receives $n-f$ \msgtype{echo} messages with the same $v$, or it receives $f+1$ \msgtype{ready} messages with the same $v$. \label{rb-step}
    
    \item Finally, when a replica receives $n-f$ \msgtype{ready} messages with the same $v$, it delivers $v$.
    
\end{enumerate}

In step~\ref{rb-step}, for a value $v$, a replica broadcasts the \msgtype{ready} message only once. For example, in \ref{fig:rbc-example}, after receiving three \msgtype{echo} messages, replica $R_1$ broadcasts a \msgtype{ready} message; it does not broadcast twice at the ready phase. In contrast, $R_3$ fails to collect $n-f$ \msgtype{echo} messages, so it must broadcast a \msgtype{ready} message when it receives $f+1$ \msgtype{ready} from the other replicas.

Bracha's RBC terminates at the end of its last phase. This guarantees that if the proposer $p$ is correct, then all correct processes deliver $p$'s proposed value, or if $p$ is faulty, then either all correct processes deliver the same value or none of them deliver $p$'s proposed value.

\subsubsection{Binary Byzantine agreement}\label{sec:ba}

\begin{figure*}[h]

    \begin{subfigure}{0.47\textwidth}
        \centering
        \includegraphics[width=\linewidth]{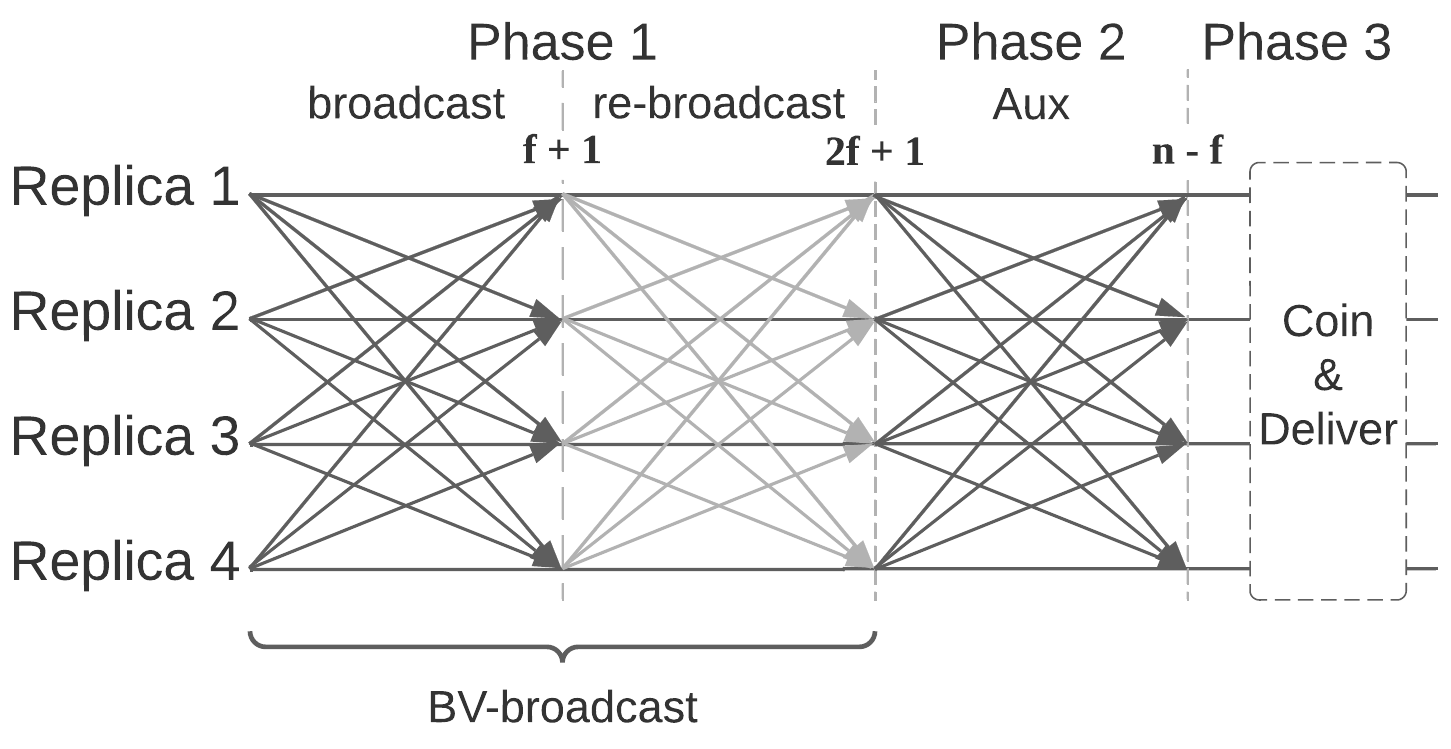}
        \caption{The message-passing under normal operation.}
        \label{fig:ba-regular}
    \end{subfigure}
    \hfill
    \begin{subfigure}{0.47\textwidth}
        \centering
        \includegraphics[width=\linewidth]{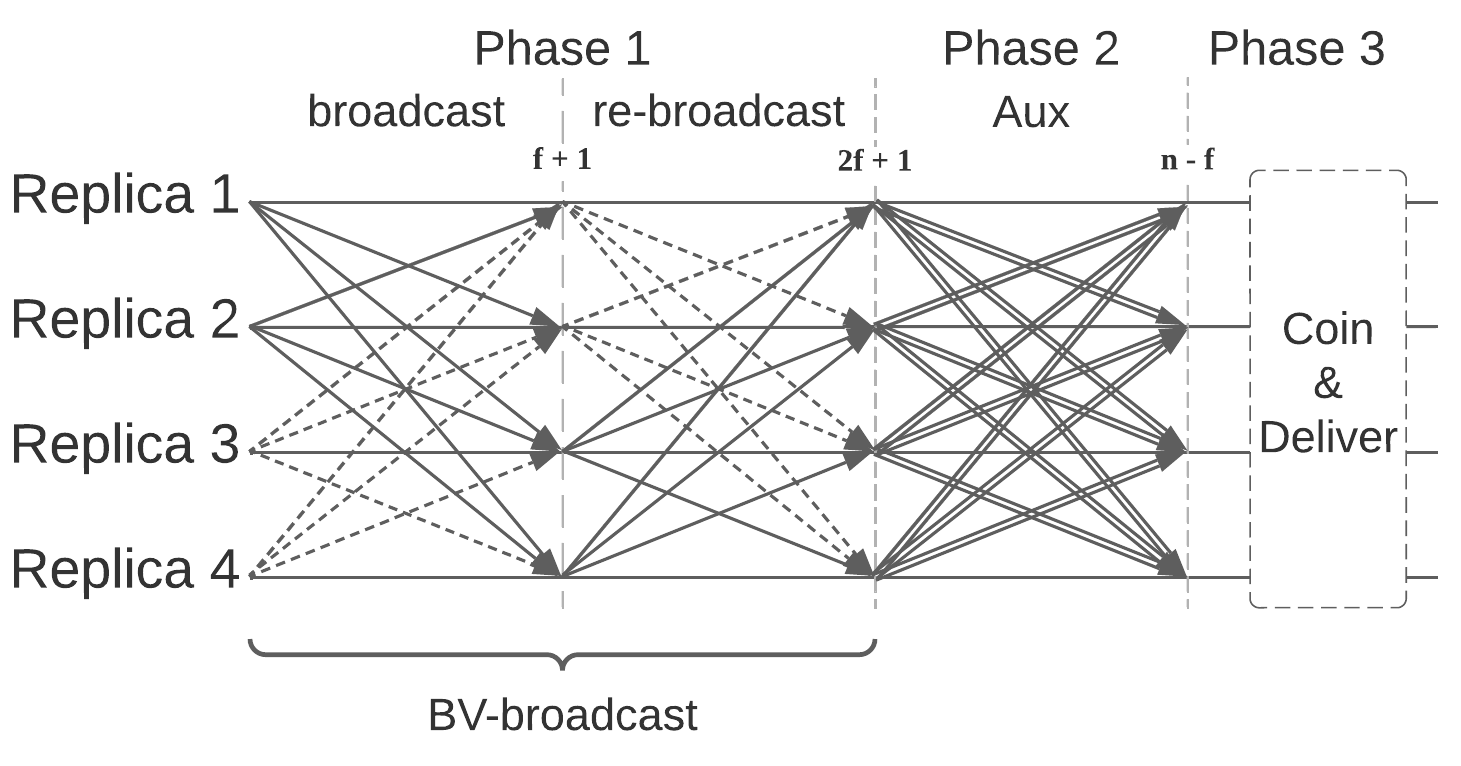}
        \caption{Resolving detected conflicts.}
        \label{fig:ba-conflict}
    \end{subfigure}
    \caption{The messaging pattern of the binary Byzantine agreement protocol where $n=4$ and $f=1$.}
    \label{fig:ba}
\end{figure*}

Similar to the network assumption of Bracha's RBC, the binary Byzantine agreement protocol~\cite{mostefaoui2014signature} operates in asynchronous networks despite failures. The protocol operates through a progression of rounds and terminates until a consensus is reached. Each round has three coordination phases aiming to deliver a value under the presence of Byzantine failures (shown in Figure \ref{fig:ba-regular}).

In Phase 1, each replica first increments its round number $r$ by one and then performs a two-step broadcast, denoted by \textit{BV-broadcast}. Each replica first proposes an estimated value ($est$), $0$ or $1$, and broadcasts the value with $r$. Then, when a replica receives the same value as the current round number $r$ from $f+1$ replicas and this value has not yet been broadcast, the replica re-broadcasts the value (gray arrows in \ref{fig:ba-regular}). In this case, if a replica receives $2f+1$ messages with the same value $v$, \textit{BV-broadcast} considers this value to be delivered, and the replica stores this value to an array $V$ (i.e., \textit{bin\_values} in the original paper), where $V$ may contain multiple values.

In Phase 2, to commit values in $V$, a replica broadcasts an \msgtype{aux} message with $V$ if its $V$ is not empty. In addition, replicas can asynchronously execute this phase as long as $V$ is not empty, so an \msgtype{aux} message can be broadcast before $V$ collects its final value.

In Phase 3, when a replica receives $n-f$ \msgtype{aux} messages, if (1) every message contains a single value $v$ and (2) $v \in V$ (i.e., $v$ was delivered by \textit{BV-broadcast} in Phase 1), the replica starts a coin-flip process to decide the final delivery of $v$ by the following three steps:

\begin{enumerate}
    \item It uses a random number generator to obtain a coin-flip value $c \in {0, 1}$.
    \item If $v = c$, it delivers $v$ and sets $est = v$ as an initial estimated value for the next round.
    \item If $v \neq c$, it sets the initial estimated value as the coin value; i.e., $est = c$ for the next round.
\end{enumerate}

Phase 3 does not involve message exchanges among replicas. Each replica decides the delivery (or abort) of the initial value of the current round and assigns the initial value for the next round.

Figure \ref{fig:ba-conflict} shows an example of replicas proposing conflicting values. In this case, replicas $R_1$ and $R_2$ propose value $0$ (shown in solid lines), whereas $R_3$ and $R_4$ propose value $1$ (in dashed lines). After Phase 1, each replica receives three messages containing value $0$ and three messages containing value $1$ after the two steps of message passing. 
For instance, $R_1$ receives two messages with $v=1$ (it receives two dashed lines and thus re-broadcasts $v=1$) and another message with $v=0$ in the first step (one solid line). As a result, every replica has two values, $0$ and $1$, in its $V$ by the end of phase 1. After Phase 2, all \msgtype{aux} messages of each replica contain two values; therefore, no replica can decide. They must flip their coins to calculate the $est$ values for a new round, so multiple rounds may be needed to reach a final agreement.  

The BA protocol obtains validity and agreement; that is, a decided value is proposed by a correct replica (validity), and no two correct processes decide different values (agreement). Since each correct replica decides at most once, with randomized replicas reaching a common coin, BA obtains termination: with the growing number of round $r$ (i.e., $r \rightarrow + \infty$), the probability of each correct replica deciding by round $r$ is $1$.
\subsection{DBFT: Efficient leaderless Byzantine consensus
and its application to blockchains}
\label{sec:dbft}
\subsubsection{System model and service properties}
Democratic Byzantine fault tolerance (DBFT) is a leaderless BFT protocol that tolerates $f$ Byzantine replicas among at least $3f + 1$ replicas~\cite{crain2017dbft, crain2018dbft}. DBFT runs on a partially synchronous network, and its messaging pattern combines RBC and a derived version of BA, namely \textit{Psync}, which uses a \textit{weak coordinator} to improve performance. DBFT has a time complexity of 6 (rounds) in the best case, given no faulty replicas and a synchronous network, and a message complexity of $O(n^3)$.

\subsubsection{The binary Byzantine consensus, Psync}  

\begin{figure*}[t]
    \begin{subfigure}{0.49\textwidth}
        \centering
        \includegraphics[width=\linewidth]{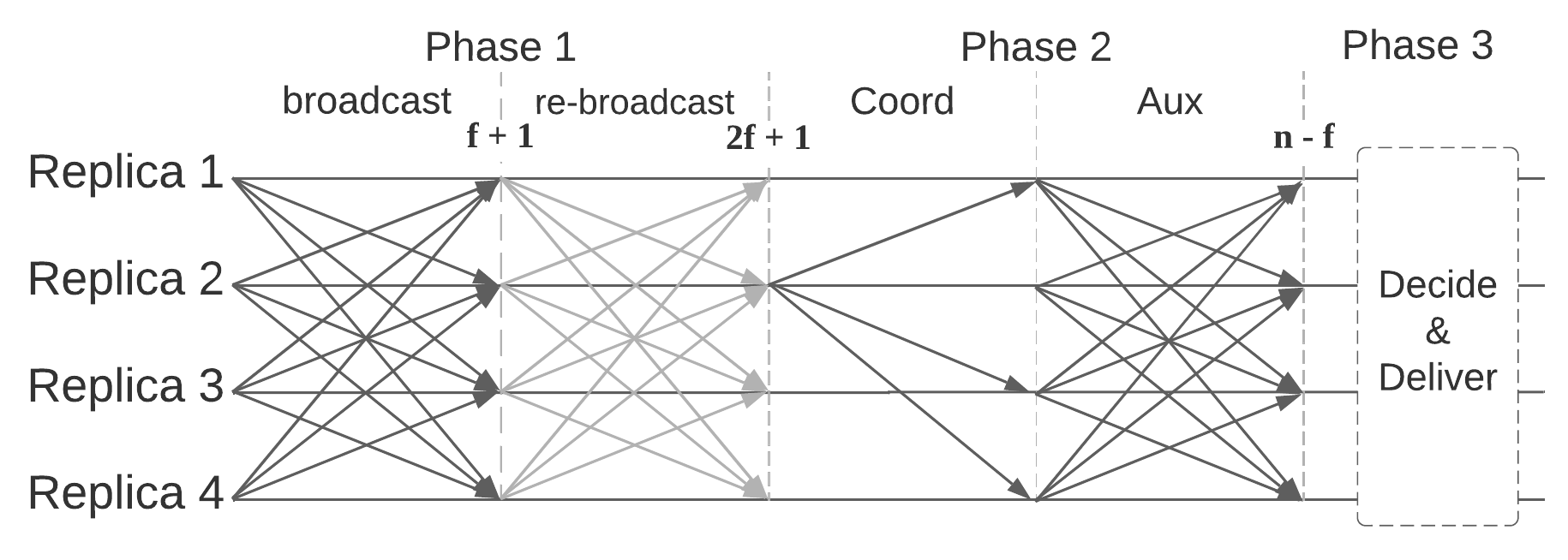}
        \caption{The message-passing under normal operation.}
        \label{fig:dbft-psync-normal}
    \end{subfigure}
    \hfill
    \begin{subfigure}{0.49\textwidth}
        \centering
        \includegraphics[width=\linewidth]{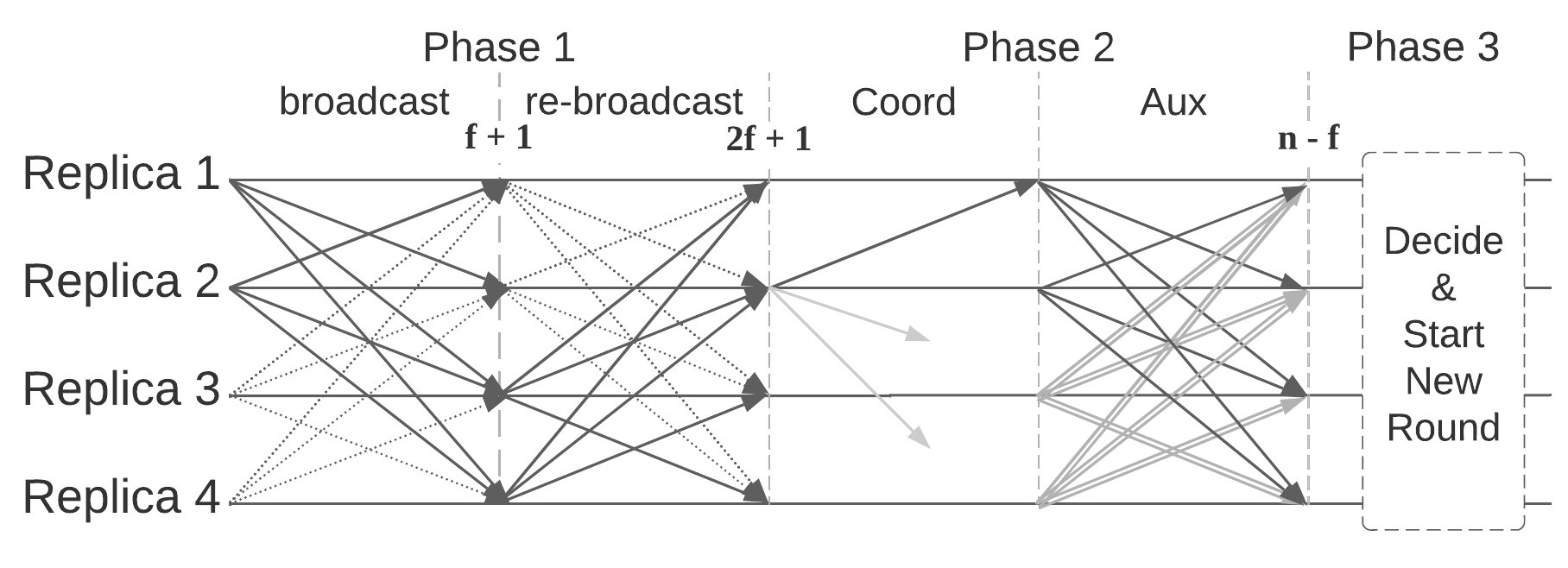}
        \caption{Resolving detected conflicts.}
        \label{fig:dbft-psync-conflict}
    \end{subfigure}
    \caption{The message-passing workflow of Psync in DBFT where $n=4$ and $f=1$.}
\end{figure*}

Psync is adapted from the BA protocol shown in Section \ref{sec:ba}. Psync made three major changes to the original protocol:

\begin{enumerate}
    \item Psync adds timeout triggers in each phase to handle partial synchronous networks.
    \item Psync adds a \textit{weak coordinator} in Phase 2 to resolve conflicts. 
    \item Psync adds a termination condition in Phase 3.
\end{enumerate}

As shown in Figure \ref{fig:dbft-psync-normal}, Psync divides Phase 2 of the BA protocol into two steps. 
A \textit{weak coordinator} $i$ is picked in a round-robin fashion such that ${i = ((r+1)\mod n)+1}$, where $r$ is the round number and $n$ is the number of all replicas.
In this example, replica $R_2$ is the weak coordinator, and it broadcasts a \msgtype{coord} message to the other replicas. When the weak coordinator has multiple values in its value set $V$, it picks the value $v$ received from the smallest ID replica and then broadcasts a \msgtype{coord} message with $v$. As a result,
Psync can resolve value conflicts to a great extent. Next, the weak coordinator and the other replicas that have received the \msgtype{coord} message within a time delay threshold broadcast an \msgtype{aux} message containing $v$. At the end of this phase, replicas wait for $n-f$ \msgtype{aux} messages to start Phase 3.

Since Psync works in partial synchrony, if a non-weak-coordinator replica does not receive any \msgtype{coord} message in time, it still broadcasts the \msgtype{aux} messages. 
Figure \ref{fig:dbft-psync-conflict} shows an example where replicas $R_1$ and $R_2$ propose value $0$ (solid lines), while $R_3$ and $R_4$ propose value $1$ (dash lines). At the end of Phase 1, each replica receives three messages containing value $0$ and three messages containing value $1$ in two rounds. In this case, both values, 0 and 1, are included in each replica's value set $V$. 
The weak coordinator $R_2$ decides to propose value $0$ because $0$ is received from $R_1$, which has the smallest ID number. 
However, the \msgtype{coord} messages sent from $R_2$ have not reached $R_3$ and $R_4$ in time. These two replicas must broadcast their \msgtype{aux} messages containing both values $0$ and $1$. In Phase 3, no replica collects sufficient \msgtype{aux} messages (${n-f = 4-1=3}$) with a single value $v$. Consequently, no replica can deliver any value, and they must start the next round to resolve this conflict.   

In Phase 3, if a replica has decided a value in round $r$, it checks whether the current value set $V$ has more than one value; i.e., some other replica has not yet decided. If so, the replica starts the next round. Each replica halts after two rounds of its decision round; i.e., if a replica has decided in round $r$, it terminates the current Psync instance in round $r+2$. 

\subsubsection{DBFT protocol based on Psync}

\begin{figure*}[t]
    \centering
    \includegraphics[width=0.85\linewidth]{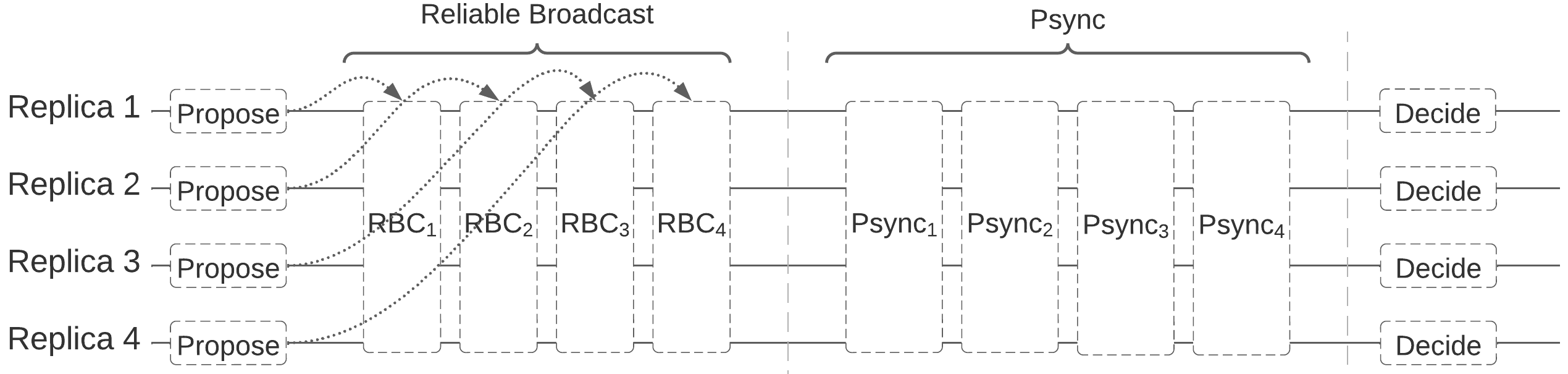}
    \caption{The complete protocol of DBFT.}
    \label{fig:dbft-complete}
\end{figure*}

Figure \ref{fig:dbft-complete} shows the complete DBFT protocol, divided into two parts.
The first part consists of Bracha's RBC~\cite{bracha1987asynchronous} instances. Each instance corresponds to a proposed value. After this instance finishes, each replica checks the validity of the proposed data, denoted by $D$, (e.g., the values proposed by all replicas). 
The second part consists of $n$ Psync instances (i.e., each replica contributes a Psync instance). This part assists all replicas in deciding whether to accept $D$. DBFT claims that these Psync instances can execute on different threads in parallel. For simplicity, we put those instances in sequential order in Figure \ref{fig:dbft-complete}. Each replica can operate those instances in two ways: 

\begin{enumerate}
    \item \textit{Fast path}. If a replica (say $j$) has RBC-delivered some data $D_i$ from replica $i$, it joins $Psync_i$, sets its Psync value set $V={1}$ in $Psync_i$, and directly starts its role in Phase 2 by skipping the double broadcast procedure in Phase 1. 
    The fast path's functionality and correctness of Phase 1 in Psync are performed and ensured by the RBC protocol (\S~\ref{sec:rbc}). Replica $j$ keeps joining such RBC-delivered Psync instances until one of the Psync instances delivers a single value $1$.  
    
    \item \textit{Regluar Path}. If a replica delivers value $1$ in a Psync instance, it joins all other Psync instances. 
    It proposes value $0$ and starts its role from Phase 1. 
    In addition, if a replica never delivers any data through RBC, it joins each Psync instance and starts from Phase 1 with a proposed value~$0$. 

\end{enumerate}

After running $n$ Psync instances, when an instance (say $Psync_i$) delivers 1, each replica picks the smallest ID $i$. Then, the data $D$ on replica $i$ are the decided data by this DBFT consensus instance.

The message complexity of each RBC or Psync instance is $O(n^2)$. Since multiple RBCs and $n$ Psync instances are needed, the total message complexity becomes $O(n^3)$. In the best case, data can be delivered with one RBC instance plus $n$ Psync instances starting from a fast path and running in parallel. In this case, the time complexity is 4 rounds (RBC) plus 2 rounds ($n$ parallel Psync with fast path), which adds up to 6 rounds. If conflicts exist, DBFT needs at least one complete round of Psync, which incurs a delay of at least 4 extra rounds.

\subsubsection{Pros and cons} 
DBFT improves the BA protocol by assigning a conflict resolver (i.e., the week coordinator). Compared with BA, Psync can directly decide a binary value in the fast path by removing the last two rounds and the random coin-flip procedure. This improvement upgrades system performance in terms of throughput and latency, though it still requires $O(f)$ delays when there are $f$ Byzantine replicas~\cite{zhang2020byzantine}. 
However, DBFT assumes a partially synchronous network to achieve consensus while some state-of-the-art leaderless protocols, such as HoenyBadgerBFT~\cite{miller2016honey} and BEAT~\cite{duan2018beat}, are able to work in asynchronous networks.
\subsection{HoneyBadgerBFT: The honey badger of BFT protocols}
\label{sec:honeybadger}

\subsubsection{System model and service properties}

HoneyBadgerBFT is a leaderless Byzantine fault-tolerant consensus protocol that tolerates $f$ Byzantine replicas in a total of $3f + 1$ replicas~\cite{miller2016honey}. 
Compared to DBFT~\cite{crain2018dbft}, which operates in partially synchronous networks, it can operate in asynchronous networks. It argues that failure detection based on timeouts requires network synchrony and tuning timeouts is difficult in practice because the system may encounter arbitrary periods of asynchrony~\cite{miller2016honey}. 
It uses threshold encryption in its protocol to defend against censorship attacks, where adversaries may refuse to broadcast or vote on certain data to prevent them from being processed. It has a time complexity of $O(\log n)$ for $n$ replicas and a message complexity of $O(|M|n + |c|n^3 \log n)$, where $M$ is the message set.

\subsubsection{Efficient reliable broadcast for large messages}\label{sec:hbrbc}

\begin{figure*}[t]
        \centering
        \includegraphics[width=0.9\linewidth]{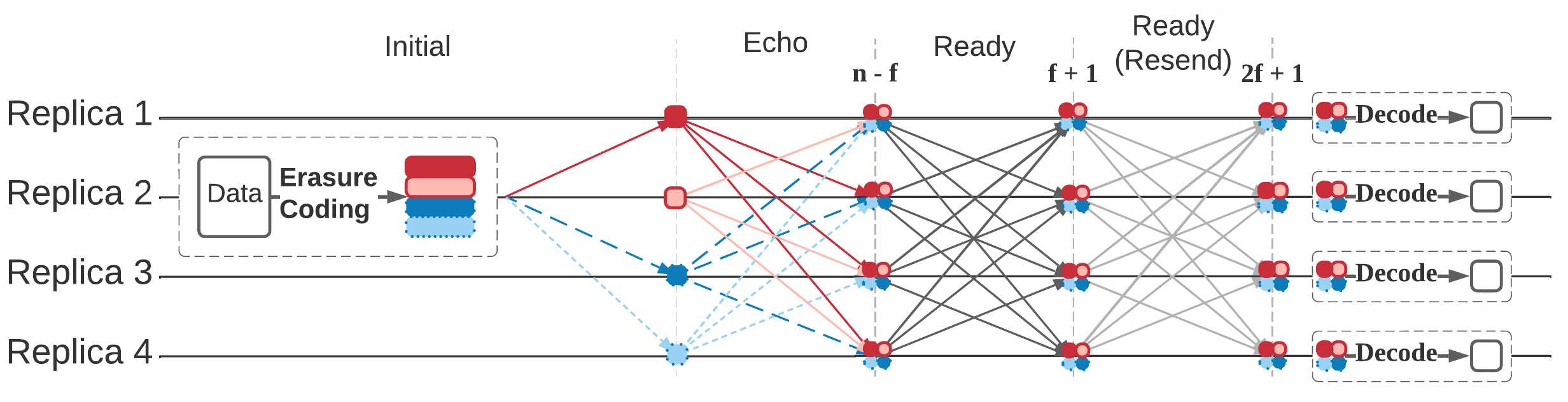}
        \caption{The reliable broadcast for large messages in HoneyBadgerBFT, with $n=4$ and $f=1$.}
        \label{fig:hb-rbc}
\end{figure*}

HoneyBadgerBFT uses batching to improve the system's throughput. When multiple RBC instances run in parallel, large messages may overload the network bandwidth and cause significant performance degradation. To reduce messaging, HoneyBadgerBFT uses erasure coding in Bracha's RBC protocol and thus reduces the communication complexity from $O(|B|n^2)$ to $O(|B|n)$, where $|B|$ is the message size of a data batch. Similar to Bracha's RBC, this protocol has four steps (shown in Figure \ref{fig:hb-rbc}).

In Step 1, the sender uses $(n-2f, n)$ Reed-Solomon codes from  \cite{wilcoxzfec} and disseminates a data batch $B$ into $n$ blocks of size $\frac{|B|}{n-2f}$. Each block contains original transactions and parity data for decoding and recovery. The storage overhead is $\frac{|B|}{n-2f} \div \frac{|B|}{n} = \frac{n}{n-2f} = \frac{3f + 1}{3f + 1 -2f} = \frac{3f+1}{f+1} < 3$, where $\frac{|B|}{n}$ is the ideal block size without any parity data. The erasure coding scheme brings two benefits:
\begin{enumerate}
    \item It reduces the size of \msgtype{initial} and \msgtype{echo} messages by a factor of $O(n)$, which further reduces the communication complexity from $O(|B|n^2)$ (in  Bracha's RBC) to $O(\frac{|B|}{n} \cdot n^2) = O(|B|n)$.  
    
    \item It requires $n-2f > f$ blocks to decode a complete batch. Thus, $f$ faulty replicas cannot produce enough blocks to pollute a correct replica's decoded data. 
\end{enumerate}

After encoding, the sender adds the Merkle tree information as a checksum $c_i$ of each encoded data block $b_i$ and then distributes each \msg{initial,$b_i$,$c_i$} to corresponding replica $i$.

In Step 2, similar to Bracha's RBC, a replica broadcasts \msg{echo,$b_i$,$c_i$} after receiving the \msgtype{initial} message. Each replica also checks the validity of the data block in each \msgtype{echo} message with its Merkle-tree-based checksum and discards any faulty ones. 

In Step 3, similar to Bracha's RBC, a replica broadcasts a \msgtype{ready} message only once if one of the following conditions holds: \textcircled{1} it receives no less than $n-f$ \msgtype{echo} messages, or \textcircled{2} it receives $f+1$ \msgtype{ready} messages with valid $h$. The replica then recalculates the Merkle root $h$ from the checksums in the messages. Once the Merkle root $h$ is valid, it broadcasts a  \msgtype{$\langle$ready, $h \rangle$} message with the Merkle root $h$ to save bandwidth.

In Step 4, when a replica receives $n-f$ \msgtype{ready} messages with valid $h$, it waits for $n-2f$ distinct \msgtype{echo} messages to collect sufficient data blocks to decode $B$; afterward, it delivers $B$.  

\subsubsection{Binary Byzantine Agreement with Cryptographic Common Coin}\label{sec:hbba} 
HoneyBadgerBFT also adapts the BA protocol \cite{mostefaoui2014signature} (introduced in~\S~\ref{sec:ba}) with two modifications: (1) a more secure coin-flip process in Phase 3, and (2) an additional termination condition in Phase 3 to prevent BA from prolonged looping.

The former process uses an $(n, f)$ threshold signature from Cachin et al. \cite{cachin2005random} to generate a common coin shared among $n=3f+1$ replicas that tolerate up to $f$ adversaries. 
This process first assigns a trusted dealer to generate a common public key ($pK$) for all replicas and $n$ secret keys ($sK_i$ for replica $i$). Each replica computes a \textit{signature share} for an input message $m$ by using their secret keys and then broadcasts the computed share. When $f+1$ shares from different replicas are collected, a replica combines these shares and produces a threshold signature, $sig_m$, that can be verified by the common public key, $pK$; $sig_m$ is valid if at least $f+1$ shares are correct. Then, the replica delivers $m$ as the output of this process. Note that $sig_m$ cannot be verified by $pK$ if some shares are faulty~\cite{shoup2000practical}. Thus, the messaging is safe with up to $f$ adversaries because polluting a delivered value requires at least $f+1$ faulty shares.

The latter modification aims to strengthen the termination of producing a common coin. HoneyBadgerBFT uses a round number ($r$) in the BA protocol as an input message (i.e., $m = r$) of the threshold signature process. Specifically, when a replica delivers a valid signature $sig_m$ (where $m = r$), it computes a coin bit such that $c = (sig_m \mod 2)$. Then, the replica checks if its agreement value set $V$ contains a value $v$ such that $v=c$. If so, it delivers the agreement set $V$.
This process requires a message complexity of $O(n^2)$ incurred by the broadcast of signature shares and a time complexity of $O(1)$ (rounds). The replica terminates the loop for obtaining a common coin (in ~\S~\ref{sec:ba}) when a new coin ($c'$) is obtained in a round such that $c' = v$.

\subsubsection{Asynchronous common subset}

HoneyBadgerBFT uses an asynchronous common subset (ACS) protocol from Ben-Or et al.~\cite{ben1994asynchronous} as the core building block. An ACS instance contains up to $n$ modified RBC instances (introduced in Section~\ref{sec:hbrbc}) and up to $n$ modified BA instances (introduced in Section~\ref{sec:hbba}) running in parallel. Each replica runs an ACS instance in two phases. 

In Phase 1, a replica $i$ starts its $RBC_i$ instance when it proposes some data $D_i$. All replicas participate in this instance if some $RBC$ from other servers starts as a proposer or a receiver.

In Phase 2, a replica joins up to $n$ BA instances to decide whether to accept some replica(s)' proposal(s). Here, if a replica delivered some data $D_j$ from $RBC_j$ (proposed by replica $j$), it sets $1$ as the input value for $BA_j$. All replicas then wait for at least $n-f$ BA instances to deliver $1$. 
After that, a replica sets $0$ to the BA instances for which it has not yet set any value and waits for all BA instances to complete. HoneyBadgerBFT expects those instances to terminate after $O(\log n)$ rounds~\cite{miller2016honey}; i.e., the time complexity is $O(\log n)$. Under HoneyBadgerBFT, some $RBC$ instance may run more slowly than its paired $BA$ instances. Thus, at the end of Phase 2, each replica waits for a paired $RBC$ and $BA = 1$ to deliver their data $D$ to itself. Finally, each replica delivers a collection of the data blocks as a final decision.

The time complexity of an ACS instance is $O(\log n)$ as it must wait for each BA to finish~\cite{ben1994asynchronous}. There are solutions with a time complexity of $O(1)$ \cite{ben2003resilient}, but HoneyBadgerBFT decides to use the current solution to achieve high throughput.  

\subsubsection{The complete protocol of HoneyBadgerBFT}  

\begin{figure*}[t]
        \centering
        \includegraphics[width=1\linewidth]{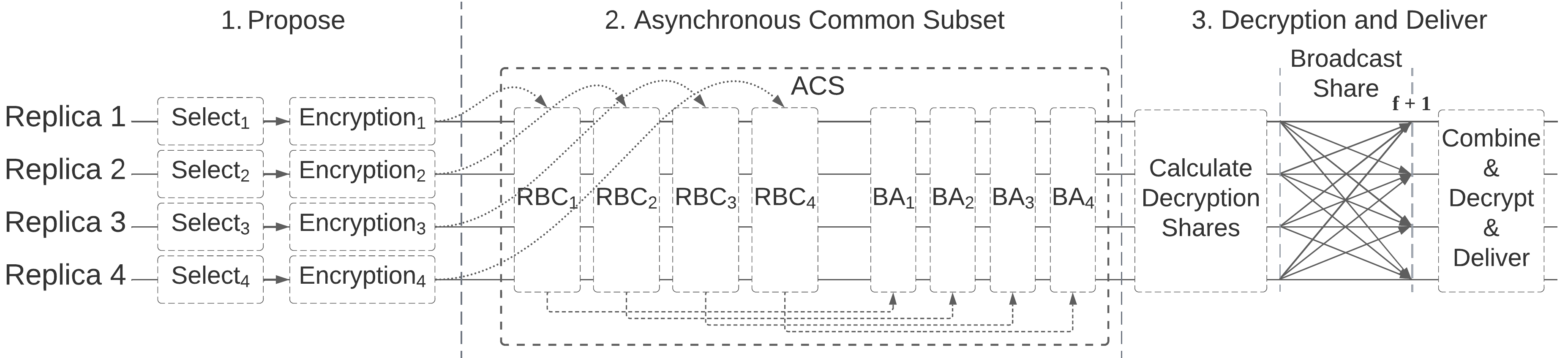}
        \caption{The complete protocol of HoneyBadgerBFT, with $n=4$ and $f=1$.}
        \label{fig:hb-full}
\end{figure*}

The complete HoneyBadgerBFT runs in three epochs (shown in \ref{fig:hb-full}) that coordinate the pace of all replicas to achieve consensus. We now show the lifetime of a transaction going through the three epochs.

In the first epoch, a replica $i$ randomly selects a batch of transactions $T_i$ from its buffer and uses a threshold encryption scheme from Baek and Zheng~\cite{baek2003simple} to encrypt $T_i$ into encrypted data $D_i$. 
This scheme works similarly to the threshold signature scheme described in~\S~\ref{sec:hbba}, where a trusted stakeholder generates a common public key and $n$ individual private keys distributed to each replica; each replica can encrypt data, produce a share, and decrypt the share with $f+1$ decryption shares.

In the second epoch, a replica $i$ uses its encrypted data $D_i$ as the input of an ACS instance. After running $n$ $RBC$ instances and $BA$ instances, the replica acquires a collection of encrypted data blocks $\{D_x | x \in \{1...n\} \land BA_x = 1\}$ delivered by this ACS instance. 

In the third epoch, a replica decrypts each acquired $D_x$ to obtain the original transaction set $T_x$. Specifically, each replica broadcasts its decryption share of $D_x$, denoted as $Share(D_x, i)$, to the other replicas. Then, if a replica collects decryption shares of a data block $D_x$ from $f+1$ different replicas, it retrieves the original data $T_x$. Once a replica retrieves the original $T_x$, it delivers transactions in $T_x$ and deletes them from its buffer. 

\subsubsection{Messaging complexity and batch sizes}

HoneyBadgerBFT utilizes many protocols to achieve consensus without a primary, which makes it tricky to calculate its messaging complexity. We analyze the messaging complexity of a consensus process that commits a set of $M$ transactions among $n$ replicas.

\begin{enumerate}
    \item The RBC. Each replica selects a subbatch of $B = \frac{M}{n}$ transactions as the input of their corresponding RBC instance. Note that each instance has a communication complexity of $O(|B|n)$ (previously analyzed in Section \ref{sec:hbrbc}), and there are $n$  RBC instances in total (shown in Figure \ref{fig:hb-full}). Therefore, the messaging complexity is $O(n \cdot |B|n) = O(|M|n)$. 
    
    \item The BA. An instance of BA requires $O(n^2)$ messages for each round, and each instance requires $O(\log n)$ rounds to guarantee termination. 
    Assume the message size in this instance is $|c|$ (the relation of $c$ and $M$ is precisely discussed in the paragraph below), each instance has a messaging complexity of $O(|c|n^2 \log n)$, and there are $n$ instances, which results in a total messaging complexity of $O(|c|n^3 \log n)$. 
    
    \item The decryption share. Each replica $i$ sends its decryption share $Share(D_x, i)$ of data blocks $D_x$ to every other replica, and it may receive $n$ blocks as a replica sends $O(n^2)$ decryption share messages. Assuming the share size is $|s|$, a replica requires a messaging complexity of $O(|s|n^2)$. With a total of $n$ replicas, the messaging complexity becomes $O(|s|n^3)$. 
    
\end{enumerate}

By summing up the above three complexities, HoneyBadgerBFT has a messaging complexity of $O(|M|n + |c|n^3 \log n +|s|n^3)$, though it claims that the decryption share size $|s|$ is much smaller than the original data block. Since $O(|s|n^3) \ll O(|c|n^3\log n)$, $O(|s|n^3)$ becomes negligible. Therefore, the messaging complexity is on the order of $O(|M|n + |c|n^3 \log n)$. In addition, to eliminate the additional overhead $O(|c|n^3 \log n)$ of $n$ replicas (i.e., $O(|c|n^2 \log n)$ overhead on each replica), HoneyBadgerBFT sets the global batch size $M = \Omega(|c|n^2 \log n)$ to achieve the best throughput performance~\cite{miller2016honey}. 

\subsubsection{Pros and cons}

HoneyBadgerBFT is able to achieve high throughput over tens of thousands of transactions per second among a hundred replicas. It claims to be the first practical solution that guarantees liveness in asynchronous networks. It also provides several encryption mechanisms to prevent censorship attacks.  

However, HoneyBadgerBFT sacrifices latency to achieve high throughput, as shown in~\cite{duan2018beat}. The latency is relatively high ($>10$s) when $n>48$~\cite{miller2016honey}, which is undesirable for applications that require low-latency consensus. The erasure coding library \cite{wilcoxzfec} of HoneyBadgerBFT also puts an upper bound of $n$ to $128$~\cite{duan2018beat}.

\subsection{BEAT: Asynchronous BFT made practical}
\label{sec:beats}

\subsubsection{System model and service properties}
BEAT~\cite{duan2018beat} consists of five leaderless asynchronous BFT protocols (i.e., BEAT0 to BEAT4), each of which addresses different goals such as throughput, latency, bandwidth usage, and BFT storage. These protocols draw inspiration from HoneyBadgerBFT~\cite{miller2016honey} and make improvements to achieve higher performance.

\subsubsection{BEAT0}

BEAT0 makes three major improvements compared to HoneyBadgerBFT.

\begin{enumerate}
    \item It uses a \emph{labeled threshold encryption} mechanism. All encrypted data blocks $D_i$ from a replica~$R_i$ in an epoch $e$ are labelled as $(e, i)$; duplicated messages with the label $(e, i)$ are discarded. In addition, BEAT0 replaces HoneyBadgerBFT's pairing-based cryptography with \texttt{TDH2}~\cite{shoup1998securing}, which significantly reduces the encryption latency~\cite{duan2018beat}.

    \item It replaces HoneyBadgerBFT's common coin protocol by directly applying a \texttt{threshold coin clipper} method~\cite{cachin2005random}; the new method can also reduce latency in obtaining a common coin. 
    
    \item It uses a more efficient erasure coding library called \texttt{Jerasure 2.0}~\cite{plankjerasure}; the new erasure coding approach has higher performance than the \texttt{zfec}~\cite{wilcoxzfec} library used in HoneyBadgerBFT and is able to scale up the cluster size to more than 128 replicas. 
\end{enumerate}

\subsubsection{BEAT1 and BEAT2}   
BEAT0 inherits the erasure-coding-based reliable broadcast protocol used in HoneyBadgerBFT; however, this protocol incurs high latency when batch sizes are small. To improve their performance, BEAT1 and BEAT2 use the original Bracha's RBC \cite{bracha1987asynchronous} to obtain lower broadcast latency, and BEAT2 also moves the labeled threshold encryption from server ends to client ends. 
By doing so, faulty replicas cannot defer transaction processing to perform censorship attacks; however, they can target and suppress specific clients to censor client requests. 
To mitigate this problem, BEAT2 uses anonymous communication networks to hide clients' information so that correct clients are not exposed to adversaries. 
This improvement assists BEAT2 in achieving casual order preservation~\cite{cachin2001secure, duan2017secure, reiter1994securely}, a weaker consistency guarantee than linearizability~\cite{linearizability1990}. 

\subsubsection{BEAT3} \label{sec:beat3}

\begin{figure*}[t]
        \centering
    \includegraphics[width=.85\linewidth]{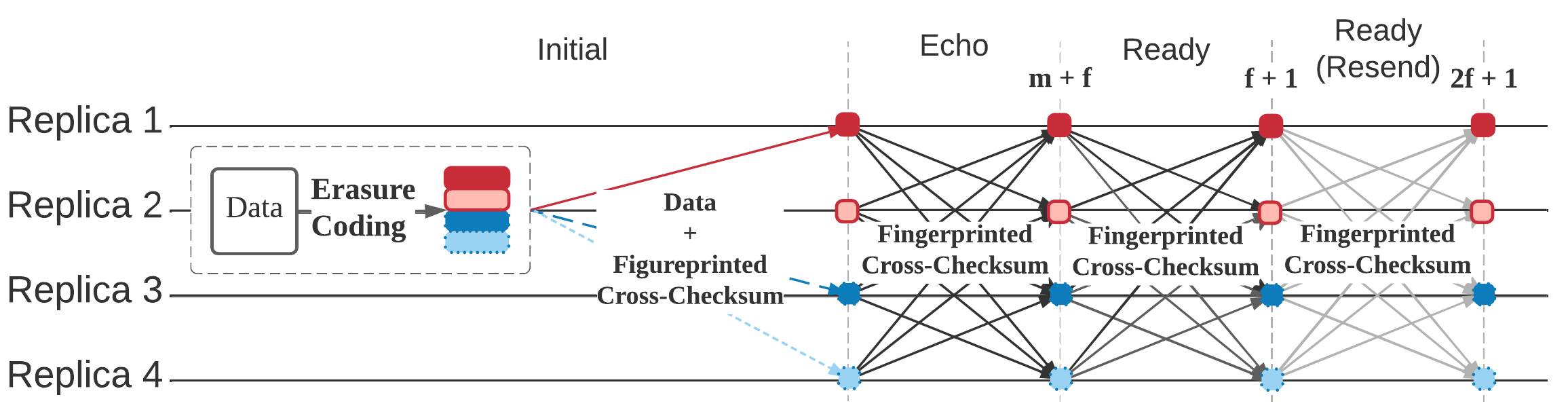}
        \caption{The reliable broadcast for large messages in BEAT3, with $n=4$, $f=1$, and $m=f+1=2$.}
        \label{fig:beat-avid-fp}
\end{figure*}

BEAT3 uses an erasure-code-based RBC protocol called \textit{AVID-FP}~\cite{hendricks2007verifying} to replace Bracha's RBC used in BEAT2. AVID-FP follows a four-phase message-passing workflow (shown in Figure~\ref{fig:beat-avid-fp}), similar to the RBC protocols introduced in~\S~\ref{sec:rbc} and~\S~\ref{sec:hbrbc}:

\begin{enumerate}
    \item A replica uses a $(n, m)$ labeled threshold encryption  to encode a data batch ($B$) to a set of fragments ($\{F_i\}$) and a single global \textit{fingerprinted cross-checksum} ($c$). It then broadcasts \msgtype{initial} messages in the form of \msg{initial, $F_i$,c} to corresponding replicas $R_i$ where $i \in \{1..n\}$. (BEAT3 sets $n = m+2f$ where $m \geq f+1$ matches the threshold).
    
    \item Each replica $R_i$ verifies the received fragment $F_i$ using cross-checksum $c$. If $F_i$ is valid, $R_i$ stores $F_i$ with $c$ and then broadcasts an \msgtype{echo} message piggybacking $c$. 
    
    \item A replica broadcasts a \msgtype{ready} message with $c$ (only once) when \textcircled{1} it has received $m+f$ \msgtype{echo} messages with $c$, or \textcircled{2} it has received $f+1$ \msgtype{ready} messages with $c$.
    
    \item After a replica receives $2f+1$ \msgtype{ready} messages with the same cross-checksum $c$, it delivers~$c$.
\end{enumerate}

Although Bracha's RBC, HoneyBadgerBFT's RBC, and AVID-FP have a similar message-passing workflow, the messaging formats are different: Bracha's RBC carries original data blocks in all messages until the protocol terminates, which involves a large message size, whereas HoneyBadgerBFT's RBC carries fragmented data blocks by the end of the second phase. In contrast, AVID-FP broadcasts fragmented data blocks only in the first phase; the rest of the phases circulate only the global cross-checksum value. Since checksums are usually much smaller than data blocks, AVID-FP is able to reduce the usage of network bandwidth to $O(|B|)+O(n|c|) = O(|B|)$ where $O(|B|)$ for the first phase and $O(n|c|)$ for the remaining phases.

The AVID-FP protocol does not guarantee that all correct replicas eventually receive the original data. Since the threshold of encryption is $f+1$, AVID-FP guarantees that $f+1$ correct replicas eventually receive $f+1$ erasure-encoded fragments of the original data, with one fragment on each replica, and all correct replicas deliver the global fingerprinted cross-checksum $c$. Nonetheless, clients can reconstruct the original data from $f+1$ fragments and verify data integrity using the cross-checksum $c$. Since a client encodes a data block $B$ and generates fragments with a checksum, the client can send a query quest to collect $f+1$ fragments paired with $c$ from $f+1$ correct replicas. Then, the client can decode the received fragments to reconstruct the original block $B$.

By using AVID-FP, BEAT3 significantly reduces the storage usage, from $O(n|M|)$ to $O(|M|)$ where $M$ is the set of all messages. The evaluation results of BEAT~\cite{duan2018beat} show that BEAT3 outperforms HoneyBadgerBFT under all performance metrics: throughput, latency, scalability, network bandwidth, and storage overhead.

\begin{table}[t]
    \centering
    
\begin{tabular}{c|cc|cc|c|c|c}
                & \multicolumn{2}{c|}{Throughput} & \multicolumn{2}{c|}{Latency} & \multirow{2}{*}{Bandwidth} & \multirow{2}{*}{Scalability} & \multirow{2}{*}{Storage} \\
                &    $b$    &     $B$   &     $b$   &    $B$    &                   &                   &                   \\ \hline
HoneybadgerBFT  &    1x      &     1x      &     1x         &    $1x$       &          $O(n|B|)$         &          Base         &       $O(n|M|)$            \\
BEAT0           &    1.1x      &    1.1x       &     0.3-0.5x      &   $\downarrow$      &          $O(n|B|)$        &        Better     &    $O(n|M|)$                 \\
BEAT1           &     1-3x      &   0.7x        &     0.3-0.5x      &   $\uparrow\uparrow$         &          $O(n^2|B|)$           &     -              &     $O(n|M|)$                \\
BEAT2           &   1.1x        &    0.7x       &     0.3-0.5x       &   $\uparrow\uparrow$         &         $O(n^2|B|)$          &       -            &       $O(n|M|)$              \\
BEAT3           &   2x        &     2x      &     0.3-0.5x       &   $\downarrow\downarrow$        &         $O(|B|)$         &     Best              &        $O(|M|)$             \\
BEAT4           &  2x         &     2x     &     0.7-1.1x      &     $\downarrow\downarrow$      &           $O(|B|)$       &          -         &        $O(|M|) + 10\%$          
\end{tabular}
    \caption{Qualitative comparison of BEAT and HoneybadgerBFT.   Notations $b$ and $B$ represent low contention (small batch size) and high contention (large batch size (e.g., $>5000$  transactions)), respectively; $\uparrow$ and $\downarrow$ represent slightly higher or lower values, and $\uparrow\uparrow$ and $\downarrow\downarrow$ represent significantly higher or lower values than the baseline; `-' means no data are presented in the paper; and $M$ is the set of all transactions that clients send.}
    \label{tab:beat-hb}
    \vspace{-2em}
\end{table}

\subsubsection{BEAT4}

This protocol replaces AVID-FP used in BEAT3 with \textit{AVID-FP-Pyramid}, which uses pyramid codes~\cite{huang2013pyramid} to reduce data reconstruction costs for read operations. Compared with MDS~\cite{macwilliams1977theory}, the conventional erasure code, pyramid codes require additional parity data but allow partial data reconstruction with less information; i.e., clients can recover specific data from tailored data fragments. This method enables a more efficient data recovery process with a lower bandwidth when clients need only partial data fragments. For example, in smart contract applications, clients often need only specific keywords to append or execute transactions; in video media services, clients may load partial fragments instead of the whole video.

\begin{wrapfigure}{r}{0.32\textwidth}
  \begin{center}
        \includegraphics[width=1\linewidth]{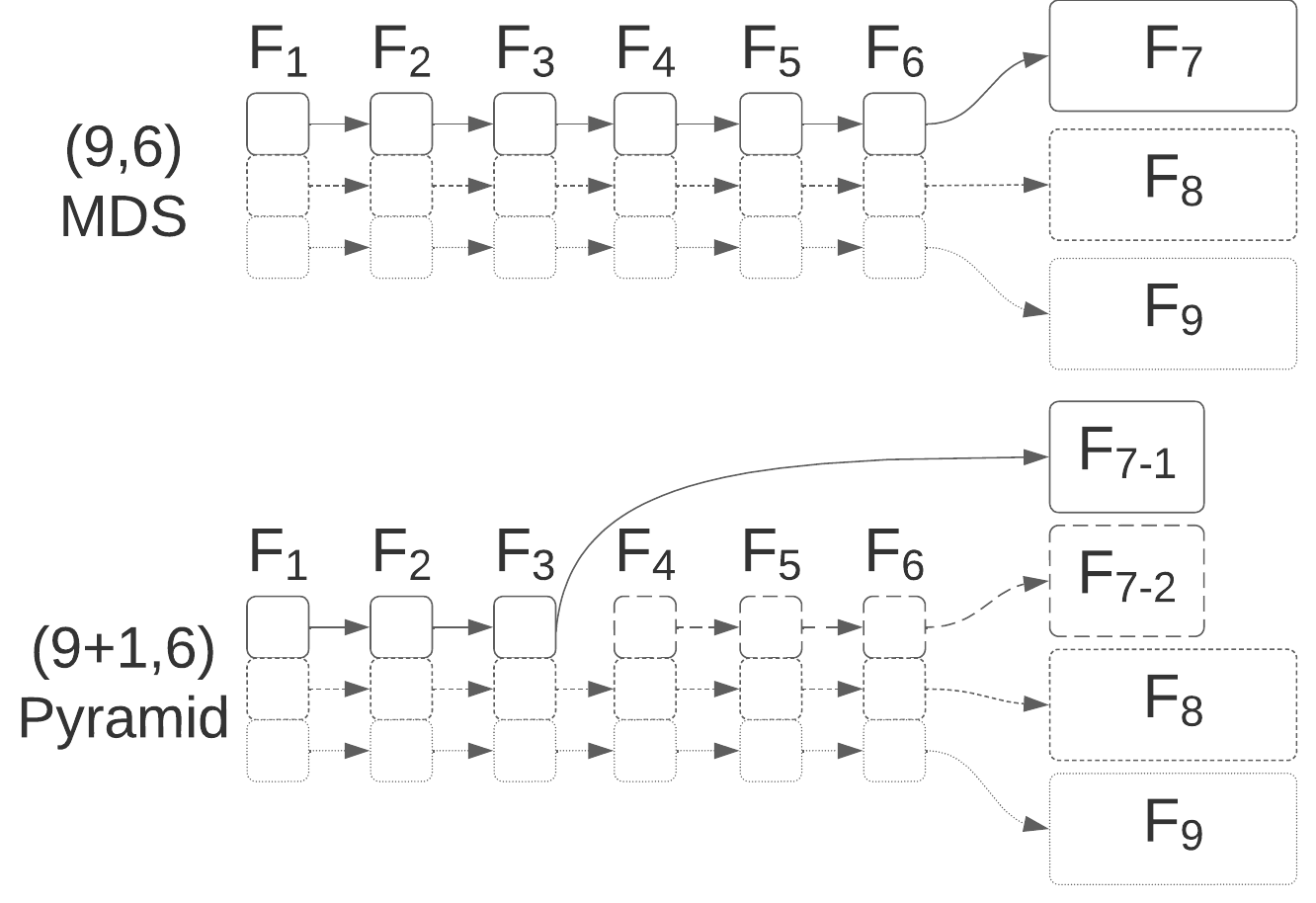}
  \end{center}
        \caption{MDS vs. pyramid codes.}
        \label{fig:beat-pyramid-code}
\end{wrapfigure}

Figure~\ref{fig:beat-pyramid-code} shows an example of the pyramid code used in BEAT4 compared to conventional MDS codes  where there are $6$ data fragments. In this example, a $(9,6)$ MDS code~\cite{macwilliams1977theory} builds three redundant parity fragments ($F_7...F_9$), whereas a $(9+1,6)=(10,6)$ pyramid code equally divides data fragments and computes two \emph{group-level} parity fragments, namely $F_{7-1}$ for $\{ F_1, F_2, F_3 \}$ and $F_{7-2}$ for $ \{ F_4, F_5, F_6 \}$, using a $(4,1)$ MDS code. Besides the two \emph{group-level} parity fragments, the $(10,6)$ pyramid code maintains two \emph{global-level} parity fragments ($F_8$ and $F_9$). In this case, if a fragment is faulty (say $F_1$), the pyramid code needs only $3$ group-level fragments ($F_2$, $F_3$ and $F_{7-1}$) to recover $F_1$ instead of using $6$ fragments in a $(9,6)$ MDS code. 

By using \textit{AVID-FP-Pyramid}, a client can directly request a specific data fragment from the corresponding replica and the checksum from $f+1$ replicas. If the fragment is faulty, the client can first try to collect group-level fragments from corresponding replicas and reconstruct the fragment. If the client fails to collect sufficient group-level fragments, it then broadcasts a request to collect global-level fragments from replicas outside the group. As a result, BEAT4 can save $50\%$ of the bandwidth in read requests, with approximately $10\%$ additional storage space~\cite{huang2013pyramid, duan2018beat}.
However, BEAT4 requires more replicas to store pyramid codes, which incurs a higher latency. For example, in BEAT's evaluation, the latency is higher than that of HoneyBadgerBFT when $f=1$, but when $f>1$, BEAT4 outperforms HoneybadgerBFT under all performance metrics.

\subsubsection{Discussion}
We qualitatively compare BEATs and HoneyBadgerBFT in Table~\ref{tab:beat-hb} in terms of throughput, latency, consumption of network bandwidth, scalability, and storage space. BEAT0, BEAT1, and BEAT2 can be applied to general state machine replication applications, while BEAT3 and BEAT4 are more suitable for BFT storage services.

\section{Future research directions}
\label{sec:future}
This paper presents a comprehensive survey of advancements in BFT consensus, covering the categories of more efficient, more robust, and more available approaches. The rise of BFT consensus applications and their studies at the intersection with other interdisciplinary areas has revealed new perspectives and challenges in the study of consensus algorithms. In this section, we identify major gaps in existing solutions and propose promising directions for future research in the field of BFT consensus algorithms, including scalability, ordering fairness, post-quantum safe consensus, BFT in machine learning, and interoperability. 

\textbf{Scalability.} 
Existing BFT algorithms perpetually navigate the delicate balance between achieving high throughput and low latency. These two facets remain ever-essential and will undoubtedly remain at the forefront of ongoing research, especially in large-scale systems. While the introduction of linear protocols (e.g., HotStuff~\cite{yin2019hotstuff} and Prosecutor~\cite{zhang2021prosecutor}) has improved scalability, it remains a practical challenge in real-world systems. For example, blockchain applications implementing BFT protocols are now expected to function across thousands of nodes, which require consensus mechanisms to maintain high throughput and low latency. Addressing scalability concerns is crucial to enable the wide adoption of BFT in increasingly large and complex networks. 

\textbf{Ordering fairness.} Fairness in transaction handling plays a pivotal role in various applications, with a profound impact on the financial sector~\cite{kelkar2020order}. Faulty leaders can unfairly handle client requests and manipulate the processing of client requests~\cite{zhang2020byzantine}. For example, a compromised leader can deliberately process transactions from its colluded clients to create front runners. These front runners exploit their advanced knowledge, for instance, by buying or selling an asset just before a large buy or sell order is executed, thereby capitalizing on the predictable price changes, which is detrimental to the fairness and transparency of the system. Thus, upholding ordering fairness will be the foundation of trust and impartiality in BFT applications.

\textbf{Post-quantum safe BFT consensus.} Current BFT consensus algorithms heavily rely on cryptography to sign messages for efficiently forming Byzantine quorums, such as PBFT~\cite{castro1999practical}, HotStuff~\cite{yin2019hotstuff}, Prosecutor~\cite{zhang2021prosecutor}, and Narwhal~\cite{danezis2022narwhal}. However, with the burgeoning development of quantum computing, traditional cryptographic methods (e.g., RSA and ECC) can be quickly solved by quantum computers. In addition to investigating post-quantum cryptography (e.g, lattice-based~\cite{peikert2014lattice}, code-based~\cite{overbeck2009code}, and multivariate polynomial cryptography~\cite{ding2005rainbow}), post-quantum safe BFT consensus algorithms can further reduce message passing to minimize the involvement of cryptography in order to defend against quantum Byzantine attacks~\cite{keidar2021all, buser2023survey}.


\textbf{Interoperability.} 
Achieving interoperability among BFT-based blockchains is becoming increasingly crucial as different blockchain networks and distributed systems emerge~\cite{han2023survey}. Researching ways to enable cross-chain consensus and interoperability will be valuable for the broader blockchain ecosystem. Developing mechanisms that facilitate seamless data and asset transfer requires consensus algorithms from both sides to coordinate agreements across platforms. Promoting interoperability will enhance overall system flexibility and utility, encouraging collaboration between different blockchain projects and fostering a more interconnected ecosystem.

\textbf{BFT consensus for machine learning.} The rapid development in distributed training has also introduced new challenges to BFT consensus. Machine learning models rely heavily on the processing of extensive datasets. In distributed training (e.g., data parallelism), each node is responsible for providing partial updates; any form of malicious or erroneous behavior during distributed training has the potential to compromise the integrity of the data, ultimately leading to inaccuracies in model updates. Thus, efficient BFT consensus will become crucial in preserving data and training integrity, underpinning the accuracy and reliability of machine learning models.

Moreover, multi-agent machine learning adds further layers of complexity to the training process~\cite{panait2005cooperative}. Multiple agents, each operating autonomously, coexist within a shared environment and make decisions in cooperative consensus~\cite{guerraoui2023byzantine}. Certain models employ consensus mechanisms where agents collectively cast their votes to determine a final decision~\cite{littman1994markov}. Thus, a high-performance and robust BFT algorithm can be used to defend against malicious agents and behavior, increasing the reliability and robustness of multi-agent machine learning models. 

These challenges are the very problems that practical BFT applications are currently facing. They represent the industry's need to build more scalable, robust, and operable BFT applications that can thrive in real-world scenarios. By addressing these critical issues, BFT research can contribute significantly to the development and adoption of secure and efficient distributed systems in both academia and industry.

\section{Conclusions}
This paper surveyed selected state-of-the-art BFT consensus algorithms that are prominent examples from academia and industry. These algorithms are categorized as efficient, robust, and available BFT algorithms. We presented a qualitative comparison of all surveyed algorithms in terms of time and message complexities. Each survey intuitively shows message-passing workflows for reaching consensus by providing diagrams that depict the process of a complete consensus instance. To improve understandability, each survey decouples the complex design and mechanism of algorithms and describes their core components of consensus following the same structure, including normal operation, view changes, robustness improvements, conflict resolutions, and a discussion of strengths and weaknesses.

\bibliographystyle{ACM-Reference-Format}
\bibliography{ref}

\end{document}